\AddToHook{package/after/caption2}{\providecommand{\captionsetup}[1]{\onelinecaptionsfalse}}
\AddToHook{package/caption2/after}{\providecommand{\captionsetup}[1]{\onelinecaptionsfalse}}
\documentclass[fleqn]{2026SCGE}
\usepackage{bm}
\usepackage{algorithm}
\usepackage{amsmath,amssymb}
\usepackage{subfigure}
\usepackage{graphicx}
\usepackage{bbm}
\usepackage{float}
\usepackage{braket}
\usepackage{anyfontsize}
\usepackage{mathtools}
\usepackage{xcolor}
\usepackage{color}
\begin{document}
\ensubject{subject}

\ArticleType{Article}
\SpecialTopic{ }
\Year{2026}
\Month{}
\Vol{}
\No{}
\DOI{}
\ArtNo{}
\ReceiveDate{}
\AcceptDate{}
\title{Helicity-Orbital Coupled Dynamics in Reciprocal Chiral Media: An Optical Dirac Framework for Photonic Rabi Oscillations}{Exact Helicity-Orbital Coupled Dynamics in Chiral Media}

\author[1]{Xu-Hui Cheng}{}
\author[1]{Long-Long Feng}{{flonglong@mail.sysu.edu.cn}}

\AuthorMark{Cheng X H, et al.}

\AuthorCitation{Cheng X H, Feng L L}

\address[1]{School of Physics and Astronomy, Sun Yat-sen University, Zhuhai 519082, China;}

\abstract{We uncover the electromagnetic origin of photonic spatial Rabi oscillations by establishing an optical Dirac framework for reciprocal chiral media. Moving beyond phenomenological coupled-mode descriptions, we show that Maxwell's equations can be recast into a \(\gamma_5\)-extended non-Hermitian Dirac-type Hamiltonian, in which the \(\gamma_5\) sector encodes electric–magnetic mismatch and helicity mixing. Breaking the spin-degenerate condition through electric–magnetic mismatch breaks electromagnetic duality and activates an intrinsic spin-orbit interaction between the helicity and orbital degrees of freedom. The resulting formulation establishes a direct and experimentally testable connection between the macroscopic constitutive parameters and spin-orbit dynamics. After projection onto the positive-eigenvalue, forward-propagating sector, the theory reduces to an effective two-level helicity–orbital model admitting an analytical solution for coherent Rabi-like oscillations. Chirality controls the helicity-dependent spectral splitting and effective detuning, whereas electric–magnetic mismatch determines the transverse coupling strength. The associated helicity flip is accompanied by a compensating change in orbital angular momentum, ensuring conservation of total angular momentum. Our results provide a unified Maxwell-level description of chirality-induced spectral splitting, intrinsic spin-orbit coupling, and coherent helicity–orbital conversion, together with a constitutive-parameter-level framework for engineering these effects in structured chiral photonic media.}

\maketitle

\begin{multicols}{2}
\section{Introduction}\label{sec1}

Light plays a central role in modern physics as a ubiquitous carrier of information in both classical and quantum regimes, in close analogy with the role of electrons in electronic systems. In recent years, rapid advances in structured light \cite{Rubinsztein-Dunlop2016,Angelsky2020,forbes2021structured} have stimulated sustained interest in the spin-orbit interaction (SOI) of light \cite{andrews2013angular,bliokh2015spin,shao2018spin,wang2018magnetic}. Optical SOI arises from the interplay between the polarization (spin) degree of freedom and the spatial structure of the electromagnetic field, including its phase, momentum, and orbital angular momentum.\Authorfootnote

\noindent 
Importantly, such coupling need not be introduced as an external perturbation: it is intrinsically encoded in the vectorial structure of Maxwell's equations. This intrinsic connection between polarization and spatial degrees of freedom gives optical SOI an operator structure closely analogous to spin-orbit coupling in relativistic quantum systems \cite{PhysRevD.90.105018,Spavieri_2015,PhysRevC.94.024304,PhysRevA.97.043840}, as well as to related spin-orbit phenomena in condensed-matter systems \cite{PhysRevLett.106.106602,PhysRevB.85.115201} and ultracold atoms \cite{li2017stripe}.

A widely used framework for describing optical SOI relies on angular momentum decomposition and geometric phases \cite{PhysRev.50.115,PhysRevLett.96.073903,chen2023spin,huang2024observation}. In classical electrodynamics, polarization directly maps to the spin degree of freedom of photons in quantum theory. Structured light carries not only spin angular momentum (SAM) but also orbital angular momentum (OAM), which can be further decomposed into extrinsic OAM (EOAM) and intrinsic OAM (IOAM). EOAM depends on the reference frame, beam centroid displacement, and propagation trajectory, and is therefore analogous to the orbital motion of classical particles. In contrast, IOAM arises solely from the helical phase structure of the wavefront and remains invariant under spatial translations of the beam axis.

The interplay between these angular-momentum degrees of freedom gives rise to a rich variety of physical effects, including the spin Hall effect of light \cite{PhysRevLett.112.113902,Ling_2017,luo2017broadband,PhysRevD.102.024075,yu2021spin,kim2023}, the quantum spin Hall effect of light \cite{bliokh2015quantum}, the orbital Hall effect \cite{Zhang_2014,PhysRevLett.123.243904,porfirev2023vectorial,choi2023observation}, and spin-orbit conversion in optical fields \cite{LUO2012864,PhysRevLett.96.163905,PhysRevLett.99.073901}. Beyond these transverse transport phenomena, the dynamical spatial evolution of structured light during propagation exhibits equally intriguing physics. Owing to intrinsic spin-orbit coupling, coherent exchange between polarization and spatial modes drives a periodic energy transfer along the propagation axis, giving rise to spatial Rabi oscillations \cite{rabiPhysRev.51.652,rabiobservationPhysRevLett.58.353,shandarova2009experimental,PhysRevLett.105.133603,zhong2019rabi,PhysRevLett.125.123201,liu2023spin}. This phenomenon establishes a direct optical analogue of driven two-level quantum dynamics.

In recent years, the concept of Rabi oscillations has transcended its origins in atomic two-level systems to become a universal dynamical framework across diverse physical platforms, ranging from the stabilization of superconducting qubits using quantum feedback \cite{vijay2012stabilizing} to acoustic Rabi oscillations between gravitational quantum states \cite{cronenberg2018acoustic}. In the optical domain, early theoretical and experimental milestones in discrete coupled optical waveguides \cite{makris2008optical} and periodic photonic lattices \cite{PhysRevLett.102.123905} successfully paved the way for manipulating the spatial evolution of light through analogue band-to-band transitions. Subsequently, research has rapidly expanded to encompass deeply subwavelength structures where full Maxwell's equations are required \cite{zhang2015rabi}, as well as the Rabi dynamics of complex structured fields. These include the spatial transitions of azimuthons carrying orbital angular momentum in weakly nonlinear waveguides \cite{jin2020rabi} and the spatial profiles of superfluorescent pulses in atomic vapors \cite{kitano2017rabi}. Moreover, these investigations have been pushed into extreme states of matter, notably the recent exploration of photonic Rabi oscillations in defective plasma photonic crystals \cite{k2th-m73q}. Very recently, spatial Rabi oscillations that simultaneously merge spin and orbital degrees of freedom have been actively investigated utilizing beam-dependent synthetic magnetic fields in artificial crystals \cite{PhysRevA.109.023522}. While these previous studies primarily focus on successfully capturing the macroscopic spatial dynamics through effective paraxial pseudospin Hamiltonians or phenomenological coupled-mode theories, our work complements these advances by shifting the emphasis to the microscopic electromagnetic origin. Specifically, rather than relying on externally engineered effective parameters, we establish an exact Maxwell-level optical Dirac framework to explicitly reveal how the intrinsic spin-orbit coupling natively emerges from the fundamental constitutive parameters of chiral media.

At a deeper level, the spin transport of light shares a common topological origin with that in condensed-matter systems and high-energy particle dynamics \cite{BERARD2006190,BARONE2010267,santos2018laser}. Optical systems therefore offer a versatile and experimentally accessible testbed for exploring spin-dependent transport and topological phenomena. Driven by advances in nanophotonics, research in spin photonics \cite{cardano2015spin,ShiDuYuan} and topological photonics \cite{lu2014topological,RevModPhys.91.015006,ma2022topological} has developed rapidly. Among various optical media, chiral materials exhibit intrinsic handedness, enabling distinct responses to left- and right-circularly polarized light. In particular, chiral metamaterials provide an engineered platform characterized by strong anisotropy and tailored electromagnetic responses \cite{wang2009chiral,wang2016optical}, supporting rich topological and transport phenomena \cite{PhysRevLett.117.057401,bhattacharjee2017topology,Shen2017TopologicalInsulators,PhysRevLett.100.013904,PhysRevA.78.033834,PhysRevLett.124.073603,PhysRevLett.114.037402,PhysRevA.93.061801,PhysRevLett.104.087401,Khanikaev2017,PhysRevResearch.3.023109}, as well as analogues in topological acoustics \cite{PhysRevLett.114.114301,Huber2022}.

While significant progress has been made in describing optical SOI using modified geometrical optics based on the eikonal approximation, second-order vector wave equations \cite{joannopoulos2008photonic}, phenomenological coupled-mode models, and device-specific Jones-matrix formalisms \cite{hosten2008observation,PhysRevLett.102.123903,hermosa2011spin,bliokh2004modified,bliokh2008geometrodynamics,PhysRevA.109.023522}, these approaches often obscure the underlying Maxwell-level Dirac structure or treat frequency dispersion purely phenomenologically. Traditional methods typically eliminate field components to form a second-order wave equation, which has long served as the standard theoretical tool for molding the flow of light in complex media \cite{joannopoulos2008photonic}. In contrast, our work adopts a first-order Dirac-like formalism. In previous studies, photonic Rabi oscillations are commonly described using an effective pseudospin Hamiltonian driven by an engineered synthetic magnetic field. 

Here, we identify its Maxwell-level electromagnetic origin within a \(\gamma_5\)-extended optical Dirac framework: electric–magnetic mismatch breaks electromagnetic duality and activates an intrinsic coupling between the helicity and orbital degrees of freedom. By generalizing the optical Dirac theory \cite{PhysRevA.106.043513,yang2025induced} to frequency-dispersive chiral metamaterials, we establish a direct connection between the macroscopic constitutive parameters and the resulting spin-orbit dynamics. The spatial Schrödinger-like evolution equation derived from the Maxwell-level Dirac Hamiltonian shows explicitly that the coupling responsible for the Rabi-like helicity–orbital oscillations is determined by the constitutive tensors rather than introduced phenomenologically. In particular, breaking the spin-degenerate condition generates the transverse helicity-mixing interaction, whereas chirality controls the helicity-dependent spectral splitting and effective detuning. This distinction provides a clear constitutive-parameter-level criterion for realizing oscillatory spin-orbit conversion in chiral photonic media.

The remainder of this paper is organized as follows. In Sec.~\ref{sec2}, we derive the dispersive optical Dirac-like eigenvalue equation in chiral media from Maxwell’s equations using a four-component spinor formalism. In Sec.~\ref{sec3}, we solve the resulting eigenvalue problem for chiral metamaterials under eikonal approximation and formulate an effective two-component spatial Schr\"odinger picture. We solve this spatial model pertubatively in momentum space, revealing frequency-tunable spatial Rabi oscillations. In Sec.~\ref{sec4}, we study SAM-OAM conversion for input vector beams carrying opposite topological charges as well as Gaussian modes, and discuss the generation of spatiotemporal optical vortices under wideband excitation. Finally, Sec.~\ref{sec5} summarizes the main conclusions.

\section{From Maxwell Equations to the Dispersive Optical Eigenvalue Equation in Chiral Media}\label{sec2}

\subsection{Helicity-Space Formulation and Emergent Optical Dirac Structure}

We consider a realistic photonic system composed of a class of uniaxial, reciprocal, and lossless chiral metamaterials. By virtue of Kramers-Kronig causality, strong magnetoelectric coupling in structured chiral media is intrinsically accompanied by frequency dispersion. Consequently, to rigorously capture this spatiotemporal memory, we formulate the electromagnetic dynamics directly in the frequency domain (or Fourier momentum space). Note that all subsequent equations are formulated in frequency space. For monochromatic Fourier field components at angular frequency $\omega$, the source-free Maxwell equations read
\begin{equation}\label{eq:MaxwellEq}
    \begin{aligned}
       & \nabla \times \mathbf{E}(\mathbf{r}) = i\omega \mathbf{B}(\mathbf{r}), \quad & \nabla \cdot \mathbf{B}(\mathbf{r}) = 0, \\
       & \nabla \times \mathbf{H}(\mathbf{r}) = -i\omega \mathbf{D}(\mathbf{r}), \quad & \nabla \cdot \mathbf{D}(\mathbf{r}) = 0.
    \end{aligned}
\end{equation}
Here, we adopt the frequency-dependent permittivity, permeability, and magnetoelectric coupling tensors for uniaxial, reciprocal, and lossless chiral metamaterials~\cite{chern2023photonic}. The electromagnetic displacement vector $\mathbf{D}(\mathbf{r})$ and magnetic induction $\mathbf{B}(\mathbf{r})$ are related to $\mathbf{E}(\mathbf{r})$ and $\mathbf{H}(\mathbf{r})$ through the dispersive constitutive relations:
\begin{equation}\label{consti}
\begin{aligned}
\mathbf{D}(\mathbf{r}) &= \boldsymbol{\epsilon} \cdot \mathbf{E}(\mathbf{r}) + i\boldsymbol{\xi} \cdot \mathbf{H}(\mathbf{r}), \\
\mathbf{B}(\mathbf{r}) &= \boldsymbol{\mu} \cdot \mathbf{H}(\mathbf{r}) - i\boldsymbol{\xi} \cdot \mathbf{E}(\mathbf{r}),
\end{aligned}
\end{equation}
where the material tensors in the principal coordinate system are diagonal matrices depending explicitly on frequency $\omega$: $\boldsymbol{\epsilon}=\text{diag}\left(\epsilon_t, \epsilon_t, \epsilon_z\right)$, $\boldsymbol{\mu}=\text{diag}\left(\mu_t, \mu_t, \mu_z\right)$, and $\boldsymbol{\xi}=\text{diag}\left(\xi_t, \xi_t, \xi_z\right)$. Inversely, the electric and magnetic fields can be expressed in terms of $\mathbf{D}(\mathbf{r})$ and $\mathbf{B}(\mathbf{r})$ via
\begin{equation}
\begin{aligned}
\mathbf{E}(\mathbf{r}) &= \boldsymbol{\varphi} \cdot \mathbf{D}(\mathbf{r}) - i\boldsymbol{\varsigma} \cdot \mathbf{B}(\mathbf{r}), \\
\mathbf{H}(\mathbf{r}) &= \boldsymbol{\vartheta} \cdot \mathbf{B}(\mathbf{r}) + i\boldsymbol{\varsigma} \cdot \mathbf{D}(\mathbf{r}),
\end{aligned}
\end{equation}
with the dispersive inverse response tensors given by
\begin{equation}\label{eq:materialtensor}
\begin{aligned}
   & \boldsymbol{\varphi}=\text{diag}\left(
                \frac{\epsilon_t^{-1}}{1-\chi_t^2},\frac{\epsilon_t^{-1}}{1-\chi_t^2},\frac{\epsilon_z^{-1}}{1-\chi_z^2}\right),\\
   & \boldsymbol{\vartheta}=\text{diag}\left(
                \frac{\mu_t^{-1}}{1-\chi_t^2},
                \frac{\mu_t^{-1}}{1-\chi_t^2},
                \frac{\mu_z^{-1}}{1-\chi_z^2}\right), \\
     &\boldsymbol{\varsigma}=\text{diag}\left(
                \frac{\chi_t}{\sqrt{\epsilon_t\mu_t}[1-\chi_t^2]},
                \frac{\chi_t}{\sqrt{\epsilon_t\mu_t}[1-\chi_t^2]},
                \frac{\chi_z}{\sqrt{\epsilon_z\mu_z}[1-\chi_z^2]}\right),
\end{aligned} 
\end{equation}
where $\chi_s=\displaystyle{\frac{\xi_s}{\sqrt{\epsilon_s\mu_s}}}$ for $s=t,z$ denotes the dimensionless frequency-dependent chirality parameter.

We generalize the optical Dirac approach developed in Ref.~\cite{PhysRevA.106.043513} to dispersive chiral media. The key idea is to reformulate Maxwell's equations in a representation that makes the internal spin structure of the electromagnetic field explicit. For a source-free electromagnetic field, we decompose the electromagnetic 6-vector \(\mathcal F=(\mathbf D,\mathbf B)^T\) into transverse and longitudinal components with respect to a given propagating direction. The transversality conditions imply that the longitudinal components are not independent but are determined by the transverse fields. The electromagnetic field can therefore be described by four independent transverse components, \(\mathcal F_\perp=(\mathbf D_\perp,\mathbf B_\perp)^T\). For each of \(\mathbf D_\perp\) and \(\mathbf B_\perp\), the two transverse components span the two-dimensional polarization space and may equivalently be expressed in the linear- or circular-polarization basis. In the following, we choose the helicity basis, in which the two circular-polarization components are eigenstates of the spin projection along the propagation direction. This reduction naturally leads to the four-component spinor-like representation of the Maxwell field used in our optical Dirac formulation.

The photon vector wave function can be constructed as follows. We work in the helicity basis spanned by the transverse vectors \(\mathbf e_\pm\), defined with respect to a reference direction \(\mathbf e_{\mathbf k}\), which we choose to coincide with the propagation direction of the incident light. Without loss of generality, we take \(\mathbf e_{\mathbf k}=\mathbf e_z\). The helicity basis satisfies
\(
(\mathbf e_{\mathbf k}\cdot\mathbf s)\mathbf e_\pm
=\pm\mathbf e_\pm ,
\) where \(\mathbf s\) denotes the spin-1 generators in the vector representation of \(\mathrm{SO}(3)\). Thus, the two transverse polarization degrees of freedom are represented by the two helicity eigenstates \(h=\pm1\). For a plane wave, these correspond to the two opposite circular-polarization states defined with respect to the propagation direction. This helicity representation provides a natural mapping from the transverse electromagnetic vector fields to the spinor-like formulation introduced below.
Projecting the displacement and magnetic induction fields onto the helicity basis, we write
\begin{equation}
\mathbf D_\perp(\mathbf r)
=
\begin{pmatrix}
D_+(\mathbf r)\\
D_-(\mathbf r)
\end{pmatrix},
\qquad
\mathbf B_\perp(\mathbf r)
=
\begin{pmatrix}
B_+(\mathbf r)\\
B_-(\mathbf r)
\end{pmatrix}.
\end{equation}
Here \(D_\pm\) and \(B_\pm\) are the amplitudes of the corresponding fields in the two helicity channels. The two entries therefore label the two physical transverse polarization degrees of freedom rather than spin-\(1/2\) states.
In analogy with the Riemann–Silberstein construction ~\cite{KELLER20051}, we introduce an alternative definition of the two-component photon wave functions in helicity space,
\begin{equation}
\boldsymbol{\Psi}^{(\pm)}(\mathbf r)
=
\mathbf D_\perp(\mathbf r)
\pm i\sigma_3\mathbf B_\perp(\mathbf r).
\end{equation}
Each \(\boldsymbol{\Psi}^{(\pm)}\) contains the two opposite-helicity components. In vacuum, these two combinations formally correspond to the two positive/negative energy branches of the first-order Maxwell evolution equation, which may equivalently be associated with the forward- and backward-propagating solutions for a prescribed positive frequency.
The complete transverse electromagnetic state can then be organized into the four-component object
\[
\boldsymbol{\Psi}_\perp(\mathbf r)
=
\begin{pmatrix}
\boldsymbol{\Psi}^{(+)}(\mathbf r)\\
\boldsymbol{\Psi}^{(-)}(\mathbf r)
\end{pmatrix}.
\]We refer to \(\boldsymbol{\Psi}_\perp\) as an effective optical bispinor. The term “bispinor” here denotes the Dirac-like organization of the four independent transverse Maxwell-field components and does not imply additional electromagnetic degrees of freedom or a fundamental spin-\(1/2\) character of the photon. This representation exposes the underlying Dirac-type algebraic structure of Maxwell's equations: the four-component state consists of two propagation branches, each containing the two physical helicity states. 

Substituting the dispersive constitutive relations into Eq.~(\ref{eq:MaxwellEq}), the magnetoelectric coupling introduces off-diagonal mixing between opposite helicity sectors. The resulting dynamics can be cast into an effective frequency-dependent optical Dirac-like eigenvalue equation for $\boldsymbol{\Psi}_\perp(\mathbf{r})$ (the detailed operator-level derivation is provided in \ref{appA}).

Specifically, given the dispersive medium constitutive tensors in Eq.~(\ref{eq:materialtensor}), the frequency-domain optical Dirac-like eigenvalue equation takes the rigorous form:
\begin{equation}\label{eq:ODE-in-chiral}
\begin{aligned}
    \omega \boldsymbol{\Psi}_{\perp}(\mathbf{r}) = \hat{H}\boldsymbol{\Psi}_{\perp}(\mathbf{r}) = &\left[\gamma_0\left(\hat{\mathrm{m}}_{+}+\gamma_5 \hat{\mathrm{m}}_{-}+\gamma_5 \gamma_3\hat{\mathrm{m}}_c\right) \right.\\&\left. +{\boldsymbol{\gamma}}\cdot\left(\hat{\mathbf{p}}_{+}+\gamma_5 \hat{\mathbf{p}}_{-}+\gamma_5\gamma_3  \hat{\mathbf{p}}_c\right)\right] \boldsymbol{\Psi}_{\perp}(\mathbf{r}),
\end{aligned} 
\end{equation}
where the dispersive mass-like and momentum-like operators are defined as
\begin{equation}\label{eq:mass_momentum_ops}
    \hat{\mathrm{m}}_{\pm} = m_{\pm}\hat{k}_z + q_{\pm}\frac{1}{\hat{k}_z} \hat{k}_{+} \hat{k}_{-}, \quad \hat{\mathbf{p}}_{\pm} = q_{\pm}\frac{1}{\hat{k}_z} \hat{\mathbf{k}}_t,
\end{equation}
and
\begin{equation}
    \hat{\mathrm{m}}_c = m_c\hat{k}_z + q_c \frac{1}{\hat{k}_z} \hat{k}_{+} \hat{k}_{-},\quad \hat{\mathbf{p}}_c = q_c \frac{1}{\hat{k}_z} \hat{\mathbf{k}}_t.
\end{equation}
Here, $\hat{k}_z = -i\partial_z$ is the longitudinal momentum operator, $\hat{\mathbf{k}}_t = (\hat{k}_+^{2}, \hat{k}_-^{2})^T$ is the transverse momentum vector, $\hat{k}_\pm = -\frac{i}{\sqrt{2}}(\partial_x \mp i\partial_y)$ are the helicity ladder operators, and the frequency-dependent scalar coefficients are strictly given by
\begin{equation}\label{eq:dispersive_coeff}
    \begin{aligned}
        &m_{\pm} = \frac{\epsilon_t^{-1}\pm\mu_t^{-1}}{2[1-\chi_t^2]},\quad &q_{\pm} &= \frac{\epsilon_z^{-1}\pm\mu_z^{-1}}{2[1-\chi_z^2]}, \\
        &m_c = \frac{\chi_t}{\sqrt{\epsilon_t\mu_t}\left[1-\chi_t^2\right]},\quad &q_c &= \frac{\chi_z}{\sqrt{\epsilon_z\mu_z}\left[1-\chi_z^2\right]}.
    \end{aligned}
\end{equation}

Equation~(\ref{eq:ODE-in-chiral}) establishes the dispersive Dirac-like representation of Maxwell's equations for a uniaxial chiral metamaterial, whose operator matrix structure is determined entirely by the optical response functions of the medium. Relative to the non-chiral static cases discussed in Refs.~\cite{PhysRevA.106.043513,yang2025induced,Wu_2022}, two additional contributions appear: the term $\gamma_0\gamma_5\gamma_3\hat{\mathrm{m}}_c$ and the term $\gamma_5\boldsymbol{\gamma}\cdot\gamma_3\hat{\mathbf{p}}_c$. Both originate from the magnetoelectric chirality tensor $\boldsymbol{\xi}$ and encode the effect of chiral symmetry breaking directly into the optical Hamiltonian $\hat{H}$.

The distinct operator coefficients in Eq.~(\ref{eq:ODE-in-chiral}) possess profound physical origins. The terms $\hat{\mathrm{m}}_+$ and $\hat{\mathbf{p}}_+$ describe the spin-independent background propagation governed by the averaged dielectric and magnetic responses of the medium. Conversely, the terms $\hat{\mathrm{m}}_-$ and $\hat{\mathbf{p}}_-$ associated with the $\gamma_5$ matrix measure the frequency-dependent mismatch between the electric and magnetic responses ($\epsilon_s \neq \mu_s, s=t,z$). They vanish when the spin-degenerate impedance-matching condition is satisfied~\cite{khanikaev2013photonic}, thereby characterizing the intrinsic breaking of electromagnetic duality and helicity degeneracy.

The coefficients $\hat{\mathrm{m}}_c$ and $\hat{\mathbf{p}}_c$ represent the genuine chiral magnetoelectric contributions. Since they are proportional to $\chi_t$ and $\chi_z$, they are absent in conventional achiral media and appear solely when the constitutive relations couple electric and magnetic fields. These chiral terms modify the optical eigenmodes by lifting the phase-velocity degeneracy between opposite helicities. The inverse response contains the factor $[1-\chi_s^2]^{-1}$, which can enhance its coefficients. Its divergence at $|\chi_s|=1$ is a singular limit of the constitutive inversion, not evidence of unbounded observable coupling. Near resonances, material loss and the validity of the propagation reduction must be assessed, as discussed in Sec.~\ref{sec:loss-validity}.

Finally, the frequency-domain Hamiltonian $\hat{H}$ in Eq.~(\ref{eq:ODE-in-chiral}) exhibits an intrinsic non-Hermitian matrix structure with respect to the standard Euclidean inner product. In the Dirac representation, $\gamma_0$, $\gamma_0\gamma_5\gamma_3$, and $\gamma_{\perp}\gamma_5$ are Hermitian matrices, whereas $\gamma_0\gamma_5$, $\gamma_{\perp}$, and $\gamma_{\perp}\gamma_5\gamma_3$ are anti-Hermitian. Consequently, the operator combinations $\gamma_0\hat{\mathrm{m}}_+$, $\gamma_0\gamma_5\gamma_3\hat{\mathrm{m}}_c$, and $\gamma_{\perp}\gamma_5\hat{\mathbf{p}}_-$ are Hermitian, while $\gamma_0\gamma_5\hat{\mathrm{m}}_-$, $\gamma_{\perp}\hat{\mathbf{p}}_+$, and $\gamma_{\perp}\gamma_5\gamma_3\hat{\mathbf{p}}_c$ are non-Hermitian. Essentially, this effective Hamiltonian non-Hermiticity does not imply optical dissipation or absorption; for reciprocal and lossless dispersive media with real material parameters, the underlying Maxwell system remains strictly conservative under the appropriate electromagnetic energy inner product.

\subsection{Symmetry of the Optical Dirac Equation in Chiral Media}

The intrinsic properties of the optical Dirac Hamiltonian are strongly constrained by the symmetries of the medium. For the system considered here, the transverse isotropy of the constitutive tensors preserves the continuous rotational symmetry about the propagation axis, namely an $\mathrm{SO}(2)\simeq U(1)$ symmetry. The corresponding conserved quantity is the z-component of the total angular momentum, $J_z=L_z+S_z$, which constrains the spin-orbit conversion processes.

We next comment on the discrete symmetry content of Eq.~(\ref{eq:ODE-in-chiral}). For a specified discrete symmetry
transformation $\Theta$, which may be unitary or antiunitary,
an individual operator contribution $\hat{O}$ is called even
or odd when
$\Theta\hat{O}\Theta^{-1}=+\hat{O}$ or
$\Theta\hat{O}\Theta^{-1}=-\hat{O}$, respectively.
This terminology applies to transformations such as spatial
inversion $\mathcal{P}$, time reversal $\mathcal{T}$, and their
combination $\mathcal{PT}$.  In particular, for the anti-linear transformation $\Theta=\mathcal{PT}$, the terms
``$\mathcal{PT}$-even'' and ``$\mathcal{PT}$-odd'' describe
the transformation properties of individual operator
contributions. They do not, by themselves, classify the full
Hamiltonian as $\mathcal{PT}$-symmetric or
anti-$\mathcal{PT}$-symmetric \cite{bender2018pt}, nor do they exclude generalized
arbitrary-phase $\mathcal{PT}$ symmetries \cite{Zheng_2021}.

Under spatial inversion $\mathcal{P}$, the electric field ${\mathbf E}$ and displacement field ${\mathbf D}$ transform as polar vectors and change sign, whereas the magnetic field ${\mathbf H}$ and magnetic induction $\mathbf B$ transform as axial vectors and remain unchanged. Since the chiral coupling connects polar and axial electromagnetic fields, the chirality parameter must transform as a pseudo-scalar under spatial inversion, $\mathcal P:\chi \rightarrow -\chi$.
Equivalently, parity maps a chiral medium of one handedness to its mirror partner with the opposite handedness. Thus, for a fixed chiral background with nonzero $\chi$, spatial inversion is not a symmetry of the system. This non-vanishing pseudo-scalar characterizes the lack of inversion and mirror symmetry of chiral media and leads to in-equivalent optical responses for the two helicities. As a result, right- and left-circularly polarized modes generally acquire different propagation constants or refractive indices.

The $\hat m_+$ and $\hat{\mathbf p}_+$ terms represent the spin-independent background propagation in the optical Dirac Hamiltonian. Their material coefficients are constructed from the symmetric combinations of the electric and magnetic responses and are therefore ordinary scalar quantities, even under both $\mathcal P$ and $\mathcal T$. Although $\hat m_+$ contains the longitudinal momentum $\hat k_z$ and hence changes sign under either spatial inversion or time reversal, the corresponding Hamiltonian term $\gamma_0\hat m_+$ changes sign in the same way under $\mathcal P$ and $\mathcal T$, and is therefore even under the combined $\mathcal{PT}$ transformation. Similarly, $\hat{\mathbf p}_+$ is invariant under simultaneous reversal of the transverse and longitudinal momenta, while the transverse Dirac matrix $\boldsymbol{\gamma}_\perp$ changes sign under both $\mathcal P$ and $\mathcal T$. As a result, the full operator $\boldsymbol{\gamma}_\perp\cdot\hat{\mathbf p}_+$ is also $\mathcal{PT}$-even. 

The spin-degeneracy-breaking terms have a distinct symmetry character. The parameter $\hat{m}_-$ itself is a parity-even scalar measuring the mismatch between the electric and magnetic responses of the medium. However, the term $\gamma_0\gamma_5 \hat{m}_-$ is parity-odd because $\gamma_5$ changes sign under spatial inversion. Similarly, $\gamma_5{\boldsymbol{\gamma}}\cdot {\mathbf p}$ is also parity-odd: the matrix structure $\gamma_5\boldsymbol{\gamma}$ transforms as an axial vector, whereas ${\mathbf p}$ is a polar momentum-like vector. In both cases, the terms are time-reversal even under the reciprocal and lossless assumptions. Therefore, these spin-degeneracy-breaking terms are $\mathcal{PT}$-odd.

Moreover, $\gamma_5{\boldsymbol{\gamma}}\cdot \hat{\mathbf p}_-$ may be interpreted as an optical analogue of a Pauli-type anomalous magnetic-moment coupling. After rewriting the Dirac-like matrix structure in the spinor basis, this term takes the schematic form
$\sim\delta\mu_a\,\sigma_3\otimes
\left(\boldsymbol{\sigma}\cdot\boldsymbol{\mathcal B}\right)$,
where $\delta\mu_a$ is an effective anomalous magnetic moment induced by the mismatch between the electric and magnetic responses of the medium, and $\boldsymbol{\mathcal B}$ denotes a pseudo-magnetic field in helicity space, which generated by the anisotropic momentum structure of the optical medium. Physically, this reflects the fact that electric-magnetic mismatch breaks the duality structure underlying helicity degeneracy, thereby allowing the two helicity sectors to couple and enabling spin-orbit conversion.

Although the chiral coefficients are proportional to the pseudo-scalar chirality parameter $\chi_s$, the full chiral terms are $\mathcal{PT}$-even. This is because the parity sign change of $\chi$ is compensated by the momentum dependence and by the axial gamma-matrix structure. In particular, $\gamma_0\gamma_5\gamma_3 \hat{m}_c$ and ${\boldsymbol{\gamma}}\cdot \gamma_5\gamma_3\hat {\mathbf p}_c$ are invariant under both $\mathcal P$ and $\mathcal T$ when the transformation properties of the chiral background are properly included. Thus, chirality breaks parity at the level of the material handedness, but the corresponding chiral terms in the optical Dirac Hamiltonian may remain $\mathcal{PT}$-even as operator combinations.

In the spin-degenerate limit, a chiral medium can still preserve electromagnetic duality symmetry. Chirality breaks spatial inversion because the chirality parameter is a pseudo-scalar, but it does not by itself imply the breaking of electric-magnetic duality. The latter is controlled by the matching between the electric and magnetic responses of the medium. When the spin-degenerate condition is satisfied, the helicity-mixing terms vanish and the two helicity sectors evolve independently. The chiral coupling then only produces different propagation constants for the two circularly polarized eigenmodes, rather than converting one helicity into the other. Thus, chirality lifts the degeneracy between opposite helicities while preserving helicity as a good quantum number. In this sense, Maxwell’s equations in a spin-degenerate chiral medium remain duality symmetric, or equivalently helicity conserving, provided that the constitutive tensors are compatible with the duality rotation.


\section{Spatial Evolution of the Dispersive Optical Eigenvalue Equation in Chiral Photonic Systems}\label{sec3}

\subsection{Frequency-Domain Eigenvalue Problem under the Eikonal Approximation}\label{Sec3:partA}

The Hamiltonian structure of the dispersive optical Dirac-like eigenvalue equation governs the dispersion relation associated with the ray system under the eikonal approximation. For a monochromatic light wave propagating along the $z$-axis, following from the Hamiltonian in Eq.(\ref{eq:ODE-in-chiral}) and retaining only the leading order of $O(k_z)$, the effective zero-order Hamiltonian of the chiral photonic system reads 
\begin{equation}\label{eq:Hamilton_eikonal}
    H_\perp^0=\begin{pmatrix}
      \hat{\mathrm{m}}_{+}- \hat{\mathrm{m}}_c\sigma_3  & \hat{\mathrm{m}}_{-} \\
      -\hat{\mathrm{m}}_{-}   & -\hat{\mathrm{m}}_{+}- \hat{\mathrm{m}}_c\sigma_3 
    \end{pmatrix},
\end{equation}
where
\begin{equation}
\hat{\mathrm{m}}_{\pm}= m_{\pm}\sigma_0\hat{k}_z,\quad \hat{\mathrm{m}}_c=m_c\sigma_0 \hat{k}_z.
\end{equation}
In a chiral photonic system, the chirality parameter of the medium introduces a mass correction in the zeroth-order Hamiltonian corresponding to a frequency-dependent pseudoscalar potential. When the chirality parameter vanishes, these additional mass terms disappear accordingly. Moreover, the off-diagonal elements of the Hamiltonian arise from the broken spin-degeneracy condition when $\boldsymbol{\epsilon}\neq \boldsymbol{\mu}$.

We now solve the eigenvalue problem associated with Eq.~(\ref{eq:Hamilton_eikonal}) for a fixed frequency $\omega$. Let 
\begin{equation}
    \boldsymbol{\Psi}_{\perp}=\left(\begin{array}{l}
\phi \\
\chi \\
\end{array}\right)e^{-iwt+ik_zz}
\end{equation}
and parameterize the dispersion relation through $\omega =\lambda k_z$, it is straightforward to write down the eigenstate equation
\begin{equation}\label{eq:eigen_eqn}
\lambda\begin{pmatrix}
\phi \\
\chi 
\end{pmatrix}=\begin{pmatrix}
m_+\sigma_0-m_c\sigma_3 & m_-\sigma_0 \\
-m_-\sigma_0 & -m_+\sigma_0-m_c\sigma_3 
\end{pmatrix}\begin{pmatrix}
\phi \\
\chi 
\end{pmatrix}.
\end{equation}

The eigenvalue problem associated with Eq.~(\ref{eq:eigen_eqn}) admits nontrivial solutions only when the determinant of the coefficient matrix vanishes. Under this condition, four frequency-dependent eigenvalues $\lambda$ can be obtained analytically as
\begin{equation}
    \begin{aligned}
        \lambda^{(+)}_{\uparrow/\downarrow} &= \sqrt{m_{+}^2-m_{-}^2}\pm m_c = \lambda_t\left[1 \pm \chi_t\right],\\
        \lambda^{(-)}_{\uparrow/\downarrow} &= -\lambda^{(+)}_{\downarrow/\uparrow} = -\lambda_t\left[1 \mp \chi_t\right],
    \end{aligned}
\end{equation}
with the baseline dispersive parameter
\begin{equation}
 \lambda_t = \frac{1}{\sqrt{\epsilon_t \mu_t}\left[1-\chi_t^2\right]}.
\end{equation}

The four eigenvalues $\lambda^{(\pm)}_{\uparrow,\downarrow}$
form positive- and negative-energy branches, labeled by the
superscripts $(+)$ and $(-)$, respectively. Introducing the
parameterization
\begin{equation}\label{para}
\sin\alpha = \frac{m_-}{m_+}
= \frac{\epsilon_t^{-1}-\mu_t^{-1}}
       {\epsilon_t^{-1}+\mu_t^{-1}},
\end{equation}
the corresponding right eigenvectors can be expressed as
\begin{equation}
u_{\uparrow}=
\begin{pmatrix}
\cos\frac{\alpha}{2}\\
0\\
-\sin\frac{\alpha}{2}\\
0
\end{pmatrix},
\qquad
u_{\downarrow}=
\begin{pmatrix}
0\\
\cos\frac{\alpha}{2}\\
0\\
-\sin\frac{\alpha}{2}
\end{pmatrix},
\end{equation}
and
\begin{equation}
v_{\uparrow}=
\begin{pmatrix}
-\sin\frac{\alpha}{2}\\
0\\
\cos\frac{\alpha}{2}\\
0
\end{pmatrix},
\qquad
v_{\downarrow}=
\begin{pmatrix}
0\\
-\sin\frac{\alpha}{2}\\
0\\
\cos\frac{\alpha}{2}
\end{pmatrix}.
\end{equation}
Here, $u$ and $v$ denote the positive- and negative-energy
branches, respectively, while $\uparrow$ and $\downarrow$
label the two spin projections along the reference
propagation axis. These vectors have unit Euclidean norm;
their biorthogonal normalization is specified below.

The four eigenvalues and corresponding eigenvectors are given by the zero-order approximation of the optical eigenvalue equation. The states $u_{\uparrow}$ and $u_{\downarrow}$ possess opposite helicities. Likewise, the helicities of the states $v_{\uparrow}$ and $v_{\downarrow}$ are also opposite to each other. A detailed inspection of the four eigenvalues further reveals a pairwise energy degeneracy. Specifically, the states $u_{\uparrow}$ and $v_{\downarrow}$ share identical energy magnitudes, while the states $u_{\downarrow}$ and $v_{\uparrow}$ possess identical energy eigenvalues. Consequently, the energy relations among these eigenstates satisfy
\begin{equation}
    \omega^{(+)}_{\uparrow}(\mathbf{k})=\omega^{(-)}_{\downarrow}({-\mathbf{k}}),\quad \omega^{(+)}_{\downarrow}(\mathbf{k})=\omega^{(-)}_{\uparrow}({-\mathbf{k}}) 
\end{equation}
where $\mathbf{k}$ and $-\mathbf{k}$ correspond to counter-propagating wave vectors. In classical electromagnetic theory, these negative-energy solutions represent backward-propagating wave modes. Because photons are their own antiparticles, these states can be understood as mirror photon modes propagating in the opposite direction.

\begin{figure}[H]
    \centering
    \includegraphics[width=0.95\linewidth]{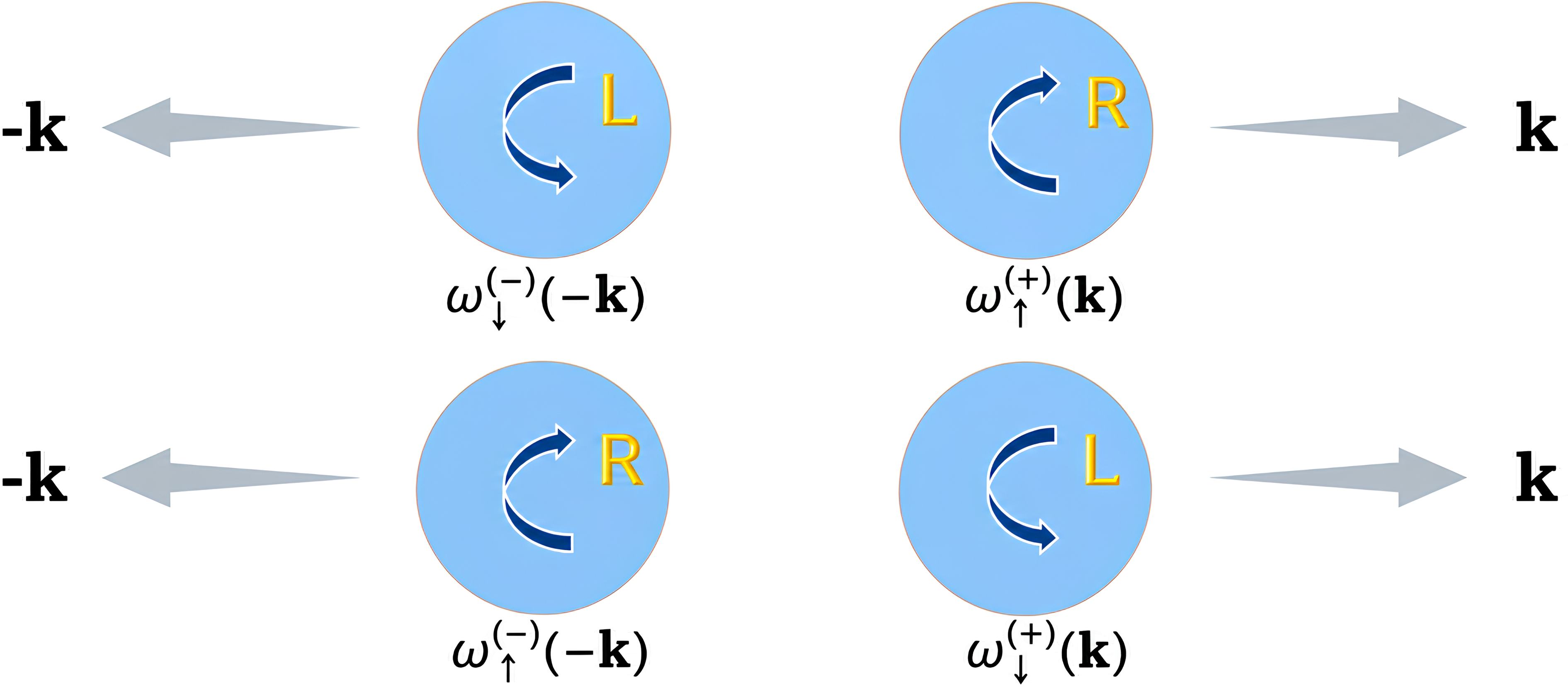}
    \caption{Schematic diagram of forward- and backward-propagating vector wave modes in helicity space. The label $\omega^{(+)}_{\uparrow/\downarrow}(\mathbf{k})$ denotes forward-propagating optical modes along wavevector $\mathbf{k}$, corresponding to positive-energy solutions with right-handed (R) and left-handed (L) helicities, respectively. Conversely, the label $\omega^{(-)}_{\downarrow/\uparrow}(-\mathbf k)$ represents negative-energy solutions propagating along $-\mathbf{k}$, which physically correspond to backward-propagating wave modes exhibiting reversed helicities (L and R, correspondingly). The helicity configurations are explicitly marked using standard R/L designations relative to the spatial propagation vector.}
    \label{figureanti}
\end{figure}

As illustrated in Fig.~\ref{figureanti}, the spectral structure naturally incorporates the reversal of the propagation direction. A forward-propagating mode with wavevector \(\mathbf{k}\) is degenerate in physical frequency with its backward-propagating counterpart at \(-\mathbf{k}\), accompanied by the corresponding reversal of helicity.  Consequently, after identifying the forward and backward solutions related by propagation reversal, the physical spectrum contains two distinct positive-frequency branches, \(\omega_{\uparrow}^{(+)}(\mathbf{k})\) and \(\omega_{\downarrow}^{(+)}(\mathbf{k})\). The four eigenmodes therefore form two frequency-degenerate pairs: \(u_{\uparrow}(\mathbf{k})\) is paired with \(v_{\downarrow}(-\mathbf{k})\), whereas \(u_{\downarrow}(\mathbf{k})\) is paired with \(v_{\uparrow}(-\mathbf{k})\). This reciprocity-related pairing constitutes a photonic Kramers-like degeneracy\cite{YANG2026170557}. 

\begin{figure*}[htpb]
    \centering
    \includegraphics[width=0.7\linewidth]{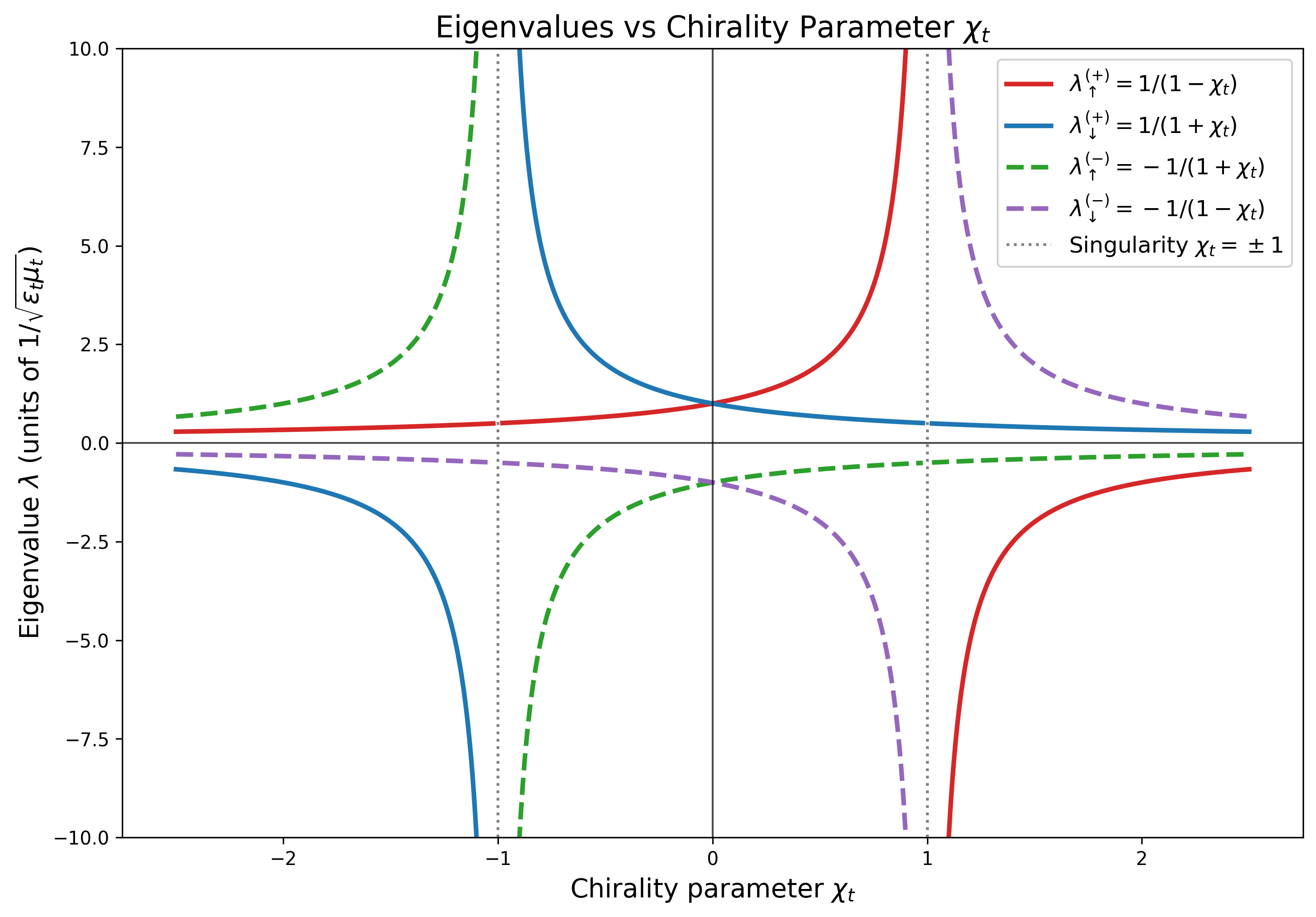}
    \caption{Eigenvalues as functions of the chirality parameter $\chi_t$, with $1/\sqrt{\epsilon_t \mu_t}$ fixed at $1.0$. The red solid line and blue solid line respectively represent the two positive-energy solutions, while the green dotted line and purple dotted line represent the two negative-energy solutions.}
    \label{eigenvalue_chi}
\end{figure*} 

Based on the four eigenvalues obtained from the optical eigenvalue equation under $\boldsymbol{\epsilon}\neq \boldsymbol{\mu}$, the chiral medium exhibits the characteristic frequency-dependent longitudinal propagation constants:
\begin{equation}\label{eq:kz_dispersion}
k_z^{0,\pm} = \frac{\omega}{\lambda_t\left[1 \pm \chi_t\right]} = \omega \sqrt{\epsilon_t \mu_t} \left[1 \mp \chi_t\right],
\end{equation}
and the corresponding phase velocities are given by
\begin{equation}\label{eq:phase-velocity}
v_p^{\pm} = \frac{\omega}{k_z^{0,\pm}} = \lambda_t\left[1 \pm \chi_t\right].
\end{equation}

In such chirality-engineered systems, the absence of spatial mirror symmetry induces a magnetoelectric coupling quantified by the chirality parameter $\xi_t$. This coupling fundamentally governs the dispersion relation by splitting the effective refractive index, which consequently modifies the phase velocities of right-handed ($v_p^{+}$) and left-handed ($v_p^{-}$) circularly polarized modes in opposite directions. As a result, the eigenvalues of the system serve directly as distinct propagation constants that dictate the spatial electromagnetic wave dynamics, ultimately generating an accumulative phase difference between the two helicities during transmission along the $z$-axis. In particular, near the regime where $\epsilon$ and $\mu$ simultaneously approach zero, one helicity may exhibit a positive refractive index while the other exhibits a negative one, as demonstrated experimentally in chiral metamaterials \cite{PhysRevLett.102.023901}.

It is important to emphasize that the \(\mathcal{PT}\)-odd character of the spin-degeneracy-breaking terms and the non-Hermiticity of the optical Hamiltonian with respect to the conventional Euclidean inner product do not imply dissipative dynamics or complex eigenfrequencies. The underlying macroscopic medium is assumed to be reciprocal and lossless, and the corresponding Maxwell dynamics therefore remains conservative with respect to the physically appropriate electromagnetic energy inner product. The apparent non-Hermiticity arises instead from the reduced helicity-spinor representation and the associated choice of inner product. In this representation, the effective Hamiltonian can be understood in a metric- or pseudo-Hermitian sense: although \(\hat H^\dagger\neq\hat H\) under the Euclidean metric, it is self-adjoint with respect to the electromagnetic metric \(\eta\), satisfying \(\hat H^\dagger\eta=\eta\hat H
\). Correspondingly, its right and left eigenmodes form a biorthogonal basis. The non-Hermitian Dirac structure therefore describes coherent mode conversion within an underlying conservative Maxwell system, rather than amplification or attenuation associated with material gain or loss.
This distinction is also explicit in the spectrum. The \(\mathcal{PT}\)-odd mismatch terms modify the eigenvalue problem through real algebraic combinations, such as \(m_+^2-m_-^2\), rather than introducing dissipative imaginary contributions. Under the passive, lossless conditions considered here, in particular \(\epsilon_t\mu_t>0\) and \(|\chi_t|<1\), the eigenvalues of the forward-propagating branches remain real. Thus, the non-Hermiticity encountered here is representation-induced and is fully compatible with conservative propagation and real eigenfrequencies.

Fig.~\ref{eigenvalue_chi} illustrates the chirality-induced energy splitting within the optical eigenvalue system. While introducing chirality ($\chi_t \neq 0$) breaks spatial inversion symmetry and lifts the initial spin degeneracy, the global spectrum remains protected by a mirror symmetry, $\lambda^{(+)}_{\uparrow}(\chi_t) = -\lambda^{(-)}_{\downarrow}(\chi_t)$. Crucially, at the singular boundaries $\chi_t = \pm 1$, the eigenvalues diverge asymptotically, signifying a near-zero-index extreme medium limit. Surpassing these critical points ($|\chi_t| > 1$) triggers a sign reversal in the propagation constant, marking a transition into an extreme physical regime characterized by backward wave propagation and reversed phase velocity.

To construct the biorthogonal basis from the eigenvectors given above, we choose the real branch $-\pi/2<\alpha<\pi/2$ and define $\mathcal{N}=(\cos\alpha)^{-1/2}$. For $h=\uparrow,\downarrow$, the normalized right eigenvectors are $|\tilde u_h^R\rangle=\mathcal{N}u_h$ and $|\tilde v_h^R\rangle=\mathcal{N}v_h$. Their left duals are defined by $\langle\tilde u_h^L|=(|\tilde u_h^R\rangle)^\dagger\gamma_0$ and $\langle\tilde v_h^L|=-(|\tilde v_h^R\rangle)^\dagger\gamma_0$, rather than by ordinary Hermitian conjugation. These definitions give
\begin{equation}\label{eq:biorthogonal-convention}
\begin{aligned}
\langle\tilde u_h^L|\tilde u_{h'}^R\rangle
&=\langle\tilde v_h^L|\tilde v_{h'}^R\rangle=\delta_{hh'},\\
\langle\tilde u_h^L|\tilde v_{h'}^R\rangle
&=\langle\tilde v_h^L|\tilde u_{h'}^R\rangle=0.
\end{aligned}
\end{equation}
The minus sign in the negative-branch left dual compensates the indefinite-metric pairing $v_h^\dagger\gamma_0v_h=-\cos\alpha$, allowing a real normalization factor for both branches. An imaginary factor obtained from $\sqrt{-\cos\alpha}$ is unnecessary; reciprocal rephasing of the left and right eigenvectors leaves their spectral projectors unchanged. This algebraic metric signature does not imply negative electromagnetic energy density or absorption and should not be identified with the physical energy metric without further derivation.

\subsection{Two-Component Helicity-Orbital Coupled Dynamics in the Spatial Schr\"odinger Picture}

To investigate the spatial evolution of optical beams carrying orbital angular momentum along the propagation axis $z$, we project the dispersive optical eigenvalue equation onto the forward-propagating positive-energy sector. For a fixed optical frequency $\omega$, the transverse optical field can be expanded in the biorthogonal basis as a linear superposition:
\begin{equation}
    \boldsymbol{\Psi}(\mathbf{r}_\perp, z)={U}_+(\mathbf{r}_\perp, z)\ket{\tilde{u}^{R}_\uparrow}e^{ik_z^{0,+}z}+{U}_-(\mathbf{r}_\perp, z)\ket{\tilde{u}^{R}_\downarrow}e^{ik_z^{0,-}z},
\end{equation}
where $k_z^{0,\pm} = k_z^0 \mp \Delta_\chi$ are the two forward-propagating longitudinal wavenumbers, with baseline wavenumber $k_z^0 = \omega \sqrt{\epsilon_t\mu_t}$ and chirality-induced splitting $\Delta_\chi = k_z^0 \chi_t$.

Introducing a two-component pseudo-spinor for the slowly varying envelope of the vector optical field in frequency space:
\begin{equation}\label{eq:redef-wavefunction}
\Phi(\mathbf{r}_\perp, z) = \begin{pmatrix} \phi_+(\mathbf{r}_\perp, z) \\ \phi_-(\mathbf{r}_\perp, z) \end{pmatrix} = \begin{pmatrix} {U}_+(\mathbf{r}_\perp, z)e^{-i\Delta_\chi z} \\ {U}_-(\mathbf{r}_\perp, z)e^{+i\Delta_\chi z} \end{pmatrix},
\end{equation}
the frequency-domain optical eigenvalue equation reduces under paraxial approximation to an effective spatial Schr\"odinger-like evolution equation:
\begin{equation}\label{eq:spatial-Schrodinger-Eq}
i\frac{\partial}{\partial z}\Phi(\mathbf{r}_\perp, z) = \hat{H}_{\text{spatial}}\Phi(\mathbf{r}_\perp, z),
\end{equation}
where the effective spatial Hamiltonian operator is given by
\begin{align}\label{eq:spatial-spinor-Hamiltonian}
    \hat{H}_{\text{spatial}} = \Delta _\chi\begin{pmatrix} 1 & 0 \\ 0 & -1 \end{pmatrix}  
 - \frac{\alpha}{2k_z^0}\begin{pmatrix} 1+\eta & 0 \\ 0 & 1-\eta \end{pmatrix}\nabla_\perp^2  + \frac{\beta}{2k_z^0}\begin{pmatrix} 0 & \hat{k}_+^2 \\ \hat{k}_-^2 & 0 \end{pmatrix}.
\end{align}
Here, the dimensionless, frequency-dependent medium parameters are strictly defined as
\begin{equation}\label{eq:def-optical-parameter}
    \begin{aligned}
        \alpha &= \frac{1}{1-\chi_z^2}\cdot\frac{1}{2}\left[\frac{\epsilon_t}{\epsilon_z}+\frac{\mu_t}{\mu_z}\right],\\
        \beta &= \frac{1}{1-\chi_z^2}\cdot\frac{1}{2}\left[\frac{\epsilon_t}{\epsilon_z}-\frac{\mu_t}{\mu_z}\right],\\
        \eta &= \chi_z \sqrt{\frac{\epsilon_t\mu_t}{\epsilon_z\mu_z}}\cdot\left\{\frac{1}{2}\left[\frac{\epsilon_t}{\epsilon_z}+\frac{\mu_t}{\mu_z}\right]\right\}^{-1}.
    \end{aligned}
\end{equation}
Apparently, $\alpha$, $\beta$, and $\eta$ characterize distinct optical properties of the dispersive metamaterial: $\alpha$ acts as an effective transverse diffraction coefficient, $\beta$ quantifies the breaking of the spin-degenerate impedance-matching condition, and $\eta$ introduces a chiral correction to the polarization-dependent diffraction. It is noteworthy that in Eq.~(\ref{eq:spatial-spinor-Hamiltonian}), the off-diagonal spin-orbit interaction term is scaled by the factor $\frac{\beta}{2k_z^0}$, which rigorously preserves the algebraic matrix structure when mapping from time-domain to spatial propagation along the $z$-axis.

The paraxial reduction applied in the above derivation is valid provided that two conditions are satisfied: (i) the propagation direction remains sufficiently close to the \(z\) axis, such that the characteristic transverse momentum is small compared with the longitudinal propagation scale; and (ii) the longitudinal field components \(E_z\) and \(H_z\) remain perturbatively small relative to the transverse fields. The latter condition further requires that the longitudinal chiral response remain sufficiently far from a resonant or singular regime, thereby avoiding anomalous enhancement of the longitudinal fields, as indicated, for example, by the factor \(1/(1-\chi_z^2)\) appearing in \(\alpha\) and \(\beta\) in Eq.~(\ref{eq:def-optical-parameter}). Under these conditions, the transverse dynamics can be consistently truncated at second order in the transverse momentum, as adopted in Eq.~(\ref{eq:spatial-spinor-Hamiltonian}). A more detailed discussion of these validity conditions is provided in ~\ref{sec:AppB}.

It is useful to decompose the spatial Hamiltonian into a diagonal
unperturbed part and an off-diagonal spin--orbit interaction:
\begin{equation}
\hat{H}_{\text{spatial}} = \hat{H}_0 + \hat{H}_1.
\end{equation}
The diagonal part is
\begin{equation}
\hat{H}_0 = \hat{\mathcal{E}}_0 \sigma_0
          + \hat{\Delta}\sigma_3,
\end{equation}
where
\begin{equation}
\hat{\mathcal{E}}_0 = -\frac{\alpha}{2k_z^0}\nabla_\perp^2, \quad
\hat{\Delta} = \Delta_\chi -\frac{\alpha}{2k_z^0}\nabla_\perp^2.
\end{equation}
Here, $\sigma_0$ is the identity matrix,
$\hat{\mathcal{E}}_0$ describes the common transverse diffraction,
and $\hat{\Delta}$ combines the longitudinal helicity detuning
with the helicity-dependent diffraction correction.
The off-diagonal interaction is
\begin{equation}\label{eq:spin-orbit-conversion}
\hat{H}_1 = \frac{\beta}{2k_z^0}
\left(
\hat{k}_+^2\sigma_+ + \hat{k}_-^2\sigma_-
\right).
\end{equation}

The scalar part $\hat{\mathcal{E}}_0$ governs the overall longitudinal phase accumulation and average transverse diffraction for both helicity components. The chirality-dependent term $\hat{\Delta}\sigma_3$ distinguishes the two helicities, generating an intrinsic spatial birefringence and chiral diffraction splitting along the propagation axis. In a chiral medium, left- and right-handed circularly polarized beams accumulate different longitudinal phases, leading to optical activity and rotation of the polarization plane during propagation. Similarly, the parameter $\eta$ induces distinct transverse spreading rates for opposite spin components.

The off-diagonal interaction $\hat{H}_1$ couples the two helicity branches via transverse orbital angular momentum. In cylindrical coordinates $(\rho, \phi, z)$, the complex transverse momentum operators read
\begin{equation}\label{eq:-OAM-operator}
\hat{k}_{\pm} = -\frac{i}{\sqrt{2}} e^{\mp i \phi}\left(\partial_\rho \pm \frac{i}{\rho} \partial_\phi\right) = -\frac{i}{\sqrt{2}} e^{\mp i \phi}\left(\partial_\rho \mp \frac{1}{\rho} L_z\right),
\end{equation}
where $L_z = -i\partial_\phi$ is the OAM operator along the $z$-direction. For OAM eigenstates satisfying $\hat{L}_z\ket{\ell}=\ell\ket{\ell}$, the operators $\hat{k}_{\pm}$ act as ladder operators that lower or raise the orbital topological charge by one unit. Meanwhile, the Pauli matrices $\sigma_{\pm} = \frac{1}{2}(\sigma_1 \pm i\sigma_2)$ drive spin-flip transitions between spin-up ($\ket{\uparrow}$) and spin-down ($\ket{\downarrow}$) states. Consequently, the operator combinations $\hat{k}_+^2\sigma_+$ and $\hat{k}_-^2\sigma_-$ describe spin flips with $\Delta S_z = \pm 2$ accompanied by simultaneous orbital transitions $\Delta L_z = \mp 2$, respectively. Therefore, each spin-orbit coupling term strictly conserves the total angular momentum $\hat{J}_z = \hat{L}_z + \hat{S}_z$, where $\hat{S}_z = \sigma_3$. One can easily verify the commutation relation $[\hat{J}_z, \hat{H}_{\text{spatial}}] = 0$, which guarantees total angular momentum conservation during spatial propagation. Finally, we emphasize that the spin-orbit interaction $\hat{H}_1$ originates entirely from the breaking of the spin-degenerate condition. In the impedance-matched limit where $\beta = 0$, the two helicity branches decouple and evolve independently in space.

\subsection{Perturbative Solution in Momentum Space: Dispersive Spatial Rabi Oscillations}

We now solve the spatial coupled-wave system governed by Eq.~(\ref{eq:spatial-Schrodinger-Eq}) in transverse momentum space. Applying the two-dimensional transverse Fourier transform at frequency $\omega$:
\begin{equation}
\Phi(\mathbf{r}_\perp, z) = \int \frac{d^2\mathbf{q}_\perp}{(2\pi)^2} \widetilde{\Phi}(\mathbf{q}_\perp, z) e^{i\mathbf{q}_\perp\cdot\mathbf{r}_\perp},
\end{equation}
the spatial Schr\"odinger equation reduces, for each transverse momentum component $\mathbf{q}_\perp = (q_x, q_y)$, to an ordinary differential equation along $z$:
\begin{equation}
i\frac{\partial}{\partial z}\widetilde{\Phi}(\mathbf{q}_\perp, z) = H(\mathbf{q}_\perp)\widetilde{\Phi}(\mathbf{q}_\perp, z).
\end{equation}
The wavenumber-space spatial Hamiltonian matrix is explicitly given by
\begin{equation}\label{eq:2-spinor-Hamq}
H(\mathbf{q}_\perp) = \mathcal{E}_0(\mathbf{q}_\perp)I + \Delta(\mathbf{q}_\perp)\sigma_3 + \Lambda_+(\mathbf{q}_\perp)\sigma_+ + \Lambda_-(\mathbf{q}_\perp)\sigma_-.
\end{equation}
where
\begin{equation}\label{eq:2-spinor-h0}
\mathcal{E}_0(\mathbf{q}_\perp) = \frac{\alpha}{2k_z^0}q_\perp^2,\quad \Delta(\mathbf{q}_\perp) = \Delta_\chi + \frac{\alpha\eta}{2k_z^0}q_\perp^2.
\end{equation}
The off-diagonal spatial spin-flip coupling matrix elements are
\begin{equation}\label{eq:2-spinor-gpm}
\Lambda_\pm(\mathbf{q}_\perp) = \frac{\beta}{2k_z^0}q_\pm^2,
\end{equation}
where $q_\pm = (q_x \mp iq_y)/\sqrt{2}$ are the complex transverse momentum coordinates, satisfying $q_+ q_- = \frac{1}{2}q_\perp^2$.

Equivalently, $H(\mathbf{q}_\perp)$ takes the familiar $2\times 2$ matrix form:
\begin{equation}\label{eq:Rab-Hamiltonian}
H(\mathbf{q}_\perp) = \begin{pmatrix} \mathcal{E}_0(\mathbf{q}_\perp) + \Delta(\mathbf{q}_\perp) & \Lambda_+(\mathbf{q}_\perp) \\ \Lambda_-(\mathbf{q}_\perp) & \mathcal{E}_0(\mathbf{q}_\perp) - \Delta(\mathbf{q}_\perp) \end{pmatrix},
\end{equation}
which is mathematically identical to a driven two-level quantum system undergoing Rabi oscillations \cite{rabiPhysRev.51.652,rabiobservationPhysRevLett.58.353,shandarova2009experimental,liu2023spin}. In our optical framework, the two states correspond to the opposite photon helicities, and the coherent driving mechanism is provided by the spatial spin-orbit interaction inherent to the anisotropic dispersive metamaterial along the propagation trajectory $z$.

When the off-diagonal perturbation is absent ($\beta=0$), the two spinor components evolve independently. The opposite helicity components experience distinct effective spatial diffraction coefficients:
\begin{equation}
D_\pm = \frac{\alpha\left[1 \pm \eta\right]}{k_z^0}.
\end{equation}

Since $H(\mathbf{q}_\perp)$ is a constant $2\times 2$ matrix for each Fourier mode at fixed frequency $\omega$, the exact formal spatial evolution solution is given by the matrix exponential:
\begin{equation}\label{eq:H-evolution-solution}
\widetilde{\Phi}(\mathbf{q}_\perp, z) = e^{-iH(\mathbf{q}_\perp)z} \widetilde{\Phi}(\mathbf{q}_\perp, 0).
\end{equation}
We define the characteristic spatial phase oscillation rate $\Omega(\mathbf{q}_\perp)$ as
\begin{equation}\label{eq:Omega}
\Omega(\mathbf{q}_\perp) = \sqrt{\Delta^2(\mathbf{q}_\perp) + \Lambda_+(\mathbf{q}_\perp)\Lambda_-(\mathbf{q}_\perp)}.
\end{equation}
Using the algebraic relation for the product of off-diagonal elements:
\begin{equation}
\Lambda_+(\mathbf{q}_\perp)\Lambda_-(\mathbf{q}_\perp) = \left[\frac{\beta}{2k_z^0}\right]^2 q_+^2 q_-^2 = \left[\frac{\beta}{2k_z^0}\right]^2 \frac{1}{4}q_\perp^4 = \left[\frac{\beta}{4k_z^0}q_\perp^2\right]^2,
\end{equation}
we obtain the exact spatial oscillation frequency:
\begin{equation}
\Omega(\mathbf{q}_\perp) = \sqrt{\Delta^2(\mathbf{q}_\perp) + \left[\frac{\beta}{4k_z^0}q_\perp^2\right]^2}.
\end{equation}
Notice that the factor of $1/4$ inside the squared coupling term arises strictly from the algebra of two-dimensional complex transverse momenta.

The spatial matrix exponential can be evaluated analytically:
\begin{equation}
e^{-iH(\mathbf{q}_\perp)z} = e^{-i\mathcal{E}_0(\mathbf{q}_\perp)z} \left[ \cos(\Omega z)I - i\frac{\sin(\Omega z)}{\Omega} \left( \Delta\sigma_3 + \boldsymbol{\Lambda}_\perp\cdot\boldsymbol{\sigma}_\perp \right) \right],
\end{equation}
where $\boldsymbol{\Lambda}_\perp\cdot\boldsymbol{\sigma}_\perp \equiv \Lambda_+\sigma_+ + \Lambda_-\sigma_-$ denotes the off-diagonal interaction in helicity space. Let the initial Fourier amplitude at $z=0$ be $\widetilde{\Phi}(\mathbf{q}_\perp, 0) = \left(\widetilde{\phi}_+(\mathbf{q}_\perp, 0), \widetilde{\phi}_-(\mathbf{q}_\perp, 0)\right)^T$. The spatial evolution of the two helicity amplitudes along $z$ is explicitly given by
\begin{align}
\widetilde{\phi}_+(\mathbf{q}_\perp, z) &= \begin{multlined}[t]
    e^{-i\mathcal{E}_0 z} \bigg[ \left(\cos\Omega z - i\frac{\Delta}{\Omega}\sin\Omega z\right)\widetilde{\phi}_+(0) \\
    - i\frac{\Lambda_+}{\Omega}\sin\Omega z\,\widetilde{\phi}_-(0) \bigg],
\end{multlined} \label{eq:phi_plus_exact} \\
\widetilde{\phi}_-(\mathbf{q}_\perp, z) &= \begin{multlined}[t]
    e^{-i\mathcal{E}_0 z} \bigg[ -i\frac{\Lambda_-}{\Omega}\sin\Omega z\,\widetilde{\phi}_+(0) \\
    + \left(\cos\Omega z + i\frac{\Delta}{\Omega}\sin\Omega z\right)\widetilde{\phi}_-(0) \bigg].
\end{multlined} \label{eq:phi_minus_exact}
\end{align}

Assuming an initial pure right-handed circularly polarized (spin-up) state, $\widetilde{\phi}_-(0) = 0$, the spatial evolution of the spin-flip inversion probability $P(z) = \langle\sigma_3\rangle = |\widetilde{\phi}_+(z)|^2 - |\widetilde{\phi}_-(z)|^2$ obeys
\begin{equation}
\begin{aligned}
P(z) =& \frac{\Delta^2(\mathbf{q}_\perp) - \Lambda_+ \Lambda_-}{\Omega^2(\mathbf{q}_\perp)}\sin^2(\Omega z) + \cos^2(\Omega z) \\=& 1 - \frac{2\Lambda_+ \Lambda_-}{\Omega^2(\mathbf{q}_\perp)}\sin^2(\Omega z).
\end{aligned}
\end{equation}
Using the trigonometric double-angle identity $\sin^2(\Omega z) = (1-\cos(2\Omega z))/2$, the macroscopic spatial Rabi oscillation frequency $\Omega_R(\mathbf{q}_\perp)$ along the propagation axis is exactly twice the phase oscillation rate:
\begin{equation}
\Omega_R(\mathbf{q}_\perp) = 2\Omega(\mathbf{q}_\perp) = 2\sqrt{\Delta^2(\mathbf{q}_\perp) + \left[\frac{\beta}{4k_z^0}q_\perp^2\right]^2}.
\end{equation}
In the phase-matched resonance limit where spatial detuning vanishes ($\Delta(\mathbf{q}_\perp)=0$), which requires eliminating both the intrinsic chirality ($\chi_t=\chi_z=0$) and differential diffraction, the spin inversion probability simplifies to a pure spatial harmonic oscillation:
\begin{equation}
P(z) = \cos\left(\Omega_R(\mathbf{q}_\perp) z\right),
\end{equation}
with the resonant spatial Rabi frequency directly governed by the dispersive anisotropy parameter:
\begin{equation}\label{eq:resonant_rabi}
\Omega_R^{\text{res}}(\mathbf{q}_\perp) = 2\sqrt{\Lambda_+ \Lambda_-} = 2\left[\frac{|\beta|}{4k_z^0}q_\perp^2\right] = \frac{|\beta|}{2k_z^0}q_\perp^2.
\end{equation}
Within the reduced model, the spatial conversion period $L_R=2\pi/\Omega_R^{\text{res}}$ depends on the dispersive coupling parameter $\beta$. Frequency tuning may shorten this period, but the experimentally observable range is limited by attenuation, detuning, and the validity of the paraxial and modal approximations; enhancement by orders of magnitude requires material-specific verification.

\section{Structured Light with OAM in Dispersive Chiral Photonic Systems}\label{sec4}

\subsection{Paraxial Spatial Modes in Chiral Media}

When the spin-degenerate impedance-matching condition is satisfied ($\beta=0$), off-diagonal orbital ladder terms such as $\hat{k}_\pm^2$ strictly vanish; consequently, the two helicity components completely decouple in space and obey independent paraxial wave equations along the $z$-axis:
\begin{equation}
i\frac{\partial {U}_\pm(\mathbf{r}_\perp, z)}{\partial z} = -\frac{\alpha_\pm}{2k_z^0}\nabla_\perp^2 {U}_\pm(\mathbf{r}_\perp, z),
\end{equation}
where the effective helicity-dependent spatial diffraction coefficients are \(\alpha_{\pm} = \alpha\left[1 \pm \eta\right]\).
Here, $U_{\pm}(r_{\perp}, z)$ denotes the amplitude envelope of the spatial electromagnetic field for each helicity component, and this spatial evolution is explicitly governed by the effective helicity-dependent macroscopic spatial diffraction coefficients, denoted by the Greek symbol $\alpha_{\pm}$.

Therefore, in the complete absence of spin-flip coupling, each spinor component can be expanded in any standard paraxial modal basis, such as Laguerre-Gaussian (LG), Hermite-Gaussian, or Bessel-Gaussian modes \cite{andrews2013angular}.

Explicitly, a monochromatic Laguerre-Gaussian modal solution at frequency $\omega$ reads
\begin{equation}\label{eq:paraxial-mode}
{U}_\pm(\mathbf{r}_\perp, z) = LG_{p_\pm, \ell_\pm}\left(\rho, \phi, z, k_\pm^{\text{eff}}\right),
\end{equation}
where $LG_{p, \ell}$ denotes a cylindrical paraxial mode with effective propagation constant $k_\pm^{\text{eff}} = k_z^0/\alpha_\pm$, given by
\begin{equation}
\begin{multlined}
LG_{p,\ell}(\rho, \phi, z, k) = \frac{a_p^\ell}{w(z)}\left[\frac{\sqrt{2}\rho}{w(z)}\right]^{|\ell|} \exp\left[-\frac{\rho^2}{w^2(z)}\right] L_p^{|\ell|}\left[\frac{2\rho^2}{w^2(z)}\right] \\
\times \exp\left[-ik\frac{\rho^2 z}{2(z_R^2+z^2)}\right] \exp(i\ell\phi) \exp\left[-i\varphi(z)\right].
\end{multlined}
\end{equation}
Here, $L_p^{|\ell|}$ are the generalized Laguerre polynomials, $p$ and $\ell$ are the radial and azimuthal quantum numbers, $N=2p+|\ell|$ is the modal order, $a_p^\ell = \sqrt{2p!/\left[\pi(p+|\ell|)!\right]}$ is the normalization factor, $w(z)=w_0\sqrt{1+(z/z_R)^2}$ is the beam width with waist $w_0$, $z_R=\frac{1}{2}k w_0^2$ is the Rayleigh range, and $\varphi(z)=(N+1)\tan^{-1}(z/z_R)$ is the Gouy phase shift.

In dispersive chiral media, the wavefront curvature and spatial beam expansion along $z$ become spin-dependent due to differential diffraction $\alpha_+ \neq \alpha_-$. To represent this difference concisely, we express the LG modes as
\begin{equation}
LG_{p,\ell}\left(\mathbf{r}, k_\pm^{\text{eff}}\right) = \ket{p, \ell} e^{\mp i\kappa z},
\end{equation}
where $\ket{p, \ell}$ represents the baseline LG transverse spatial profile. 

\subsection{Spatial Rabi-Like Oscillations for Structured Vector Light}\label{rabi}

Given the unperturbed paraxial spatial modes, we now investigate the spatial evolution governed by the full Hamiltonian in Eq.~(\ref{eq:spatial-spinor-Hamiltonian}). The off-diagonal spin-orbit interaction term $\hat{H}_1$ couples spatial modes with opposite spin polarizations. Because the spatial Hamiltonian strictly commutes with total angular momentum, $[\hat{J}_z, \hat{H}_{\text{spatial}}] = 0$, the spatial evolution preserves $J_z = L_z + S_z$. Within a two-mode truncation of a conserved angular-momentum sector, we approximate the optical state as
\begin{equation}\label{eq:two-spinor-expansion}
\Phi(\mathbf{r}_\perp, z) = C_+(z)\ket{+} + C_-(z)\ket{-},
\end{equation}
where $\ket{\pm}$ represent tensor products of spatial OAM and spin polarization states. For instance, choosing $\ket{+} = \ket{\mp 2}\otimes\ket{\uparrow/\downarrow}$ and $\ket{-} = \ket{\circ, 0}\otimes\ket{\downarrow/\uparrow}$ guarantees that both states belong to the identical conserved sector $J_z = \mp 1$. Here, $\ket{\circ, \ell} = \sum_p a_p LG_{p, \ell}(\rho, \phi, z, k)$ denotes a coherent radial superposition of LG modes carrying azimuthal charge $\ell$.

Projecting expansion~(\ref{eq:two-spinor-expansion}) onto the chosen pair of modes yields the approximate two-mode spatial coupled equations:
\begin{equation}\label{eq:Rabi-osc-eq}
i\frac{d}{dz}\begin{pmatrix} C_+(z) \\ C_-(z) \end{pmatrix} = \begin{pmatrix} \Delta_\chi & \Omega_+ \\ \Omega_- & -\Delta_\chi \end{pmatrix} \begin{pmatrix} C_+(z) \\ C_-(z) \end{pmatrix},
\end{equation}
where the spatial coupling matrix elements are evaluated via the ladder operators:
\begin{equation}\label{eq:mateix-elements}
\Omega_+ = \frac{\beta}{2k_z^0}\bra{+}\hat{k}_+^2\sigma_+\ket{-}, \quad \Omega_- = \frac{\beta}{2k_z^0}\bra{-}\hat{k}_-^2\sigma_-\ket{+} = \Omega_+^\dagger.
\end{equation}
For a complete cylindrical LG basis, the differential operators $\nabla_\pm$ act as step operators satisfying
\begin{equation}\label{eq:selection-rule}
\begin{aligned}
\nabla_{\pm}\ket{p, \pm \ell} &= k_w\left[\sqrt{p+\ell}\ket{p, \pm(\ell-1)} + \sqrt{p+1}\ket{p+1, \pm(\ell-1)}\right], \\
\nabla_{\mp}\ket{p, \pm \ell} &= k_w\left[\sqrt{p+\ell+1}\ket{p, \pm(\ell+1)} - \sqrt{p}\ket{p-1, \pm(\ell+1)}\right],
\end{aligned}
\end{equation}
where $k_w = 1/w_0$. Applying these ladder rules successively allows evaluation of the spatial coupling coefficients $\Omega_\pm$.

We examine two complementary incident configurations to illustrate dispersive spatial Rabi oscillations. First, consider an incident spin-orbit-coupled vector vortex beam at frequency $\omega$:
\begin{equation}\label{eq:nonuniform-vextex}
\ket{\mathrm{in}} = \frac{1}{\sqrt{2}} \left[ \ket{0, -2}e^{-i\phi_p(z)} \otimes \ket{\uparrow} + \ket{0, +2} e^{i\phi_p(z)}\otimes \ket{\downarrow} \right],
\end{equation}
where $\phi_p(z) = \phi_0 + \kappa z$. Here, the spin-up component carries OAM topological charge $\ell=-2$, while the spin-down component carries $\ell=+2$, thereby synthesizing a second-order vector vortex beam with a nonuniform polarization texture \cite{Stalder:96,Bomzon:02}.

\begin{figure*}[htpb]
    \centering
    \includegraphics[width=0.85\linewidth]{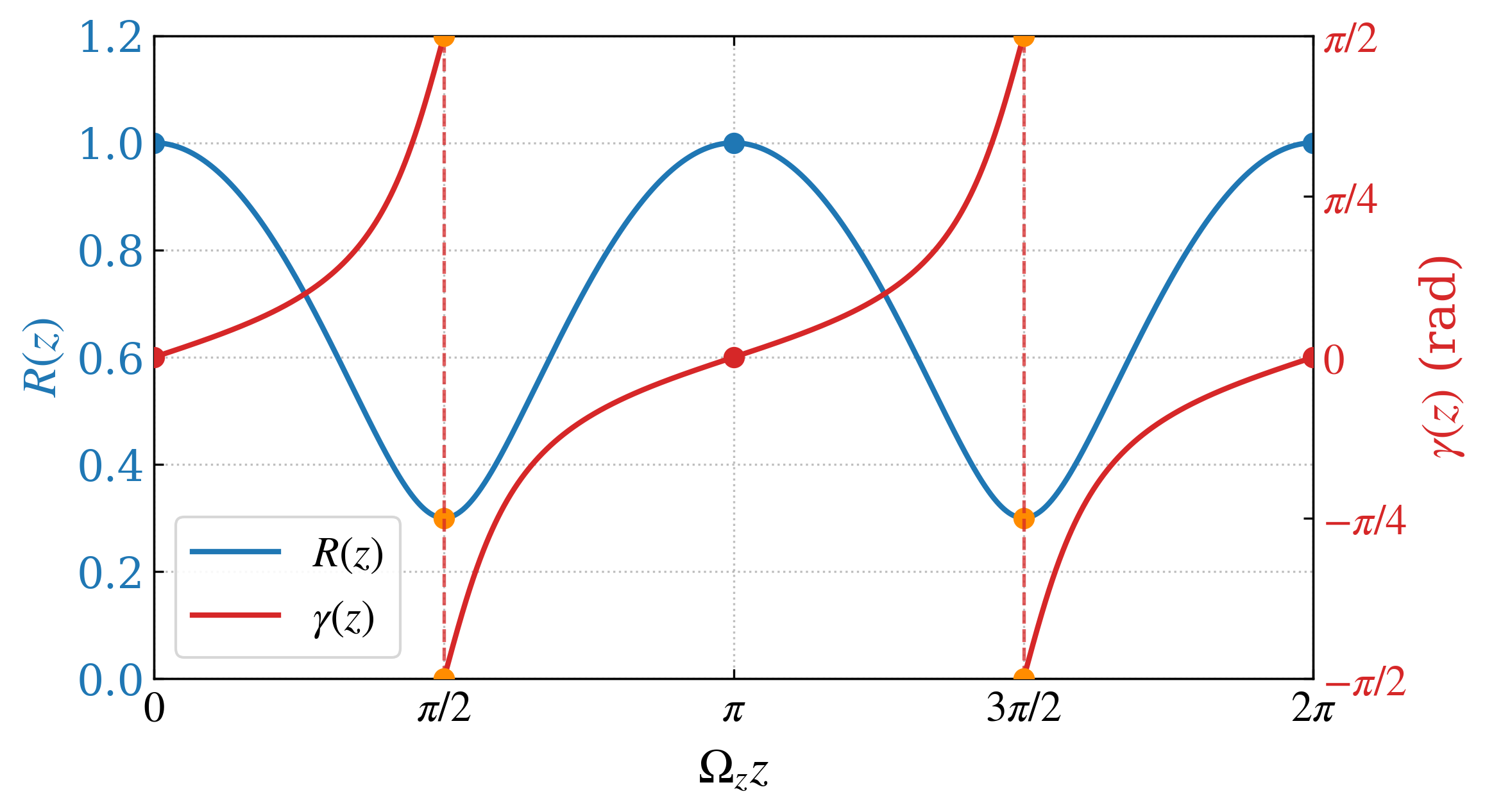}
    \caption{Spatial evolution of the amplitude factor $R(z)$ (blue curve) and relative phase $\gamma(z)$ (red curve) along the propagation axis $z$ for the intermediate-detuning regime $\delta_d=0.3$. Blue markers denote slow-evolution spatial positions ($\Omega_z z = 0, \pi, 2\pi$), where $R(z)=1$ and the spatial phase velocity reaches its minimum value $d\gamma/dz = \delta_d\Omega_z = 0.3\,\Omega_z$. Orange markers indicate fast-evolution positions ($\Omega_z z = \pi/2, 3\pi/2$), where $R(z) = \delta_d = 0.3$ and the spatial phase velocity reaches its maximum $d\gamma/dz = \Omega_z/\delta_d \simeq 3.33\,\Omega_z$. Discontinuities in $\gamma(z)$ originate from the principal branch of the arctangent function, corresponding to spatial phase jumps of $\pi$. Noticeably, the velocity ratio $\dot{\gamma}_{\max}/\dot{\gamma}_{\min} = 1/\delta_d^2 \simeq 11.1$ reveals a strongly nonuniform spatial phase evolution, which governs the nonlinear rotation dynamics of the transverse polarization texture along $z$.}
    \label{fig:R-Gamma-varing}
\end{figure*}

\begin{figure*}
    \centering
    \includegraphics[width=0.9\linewidth]{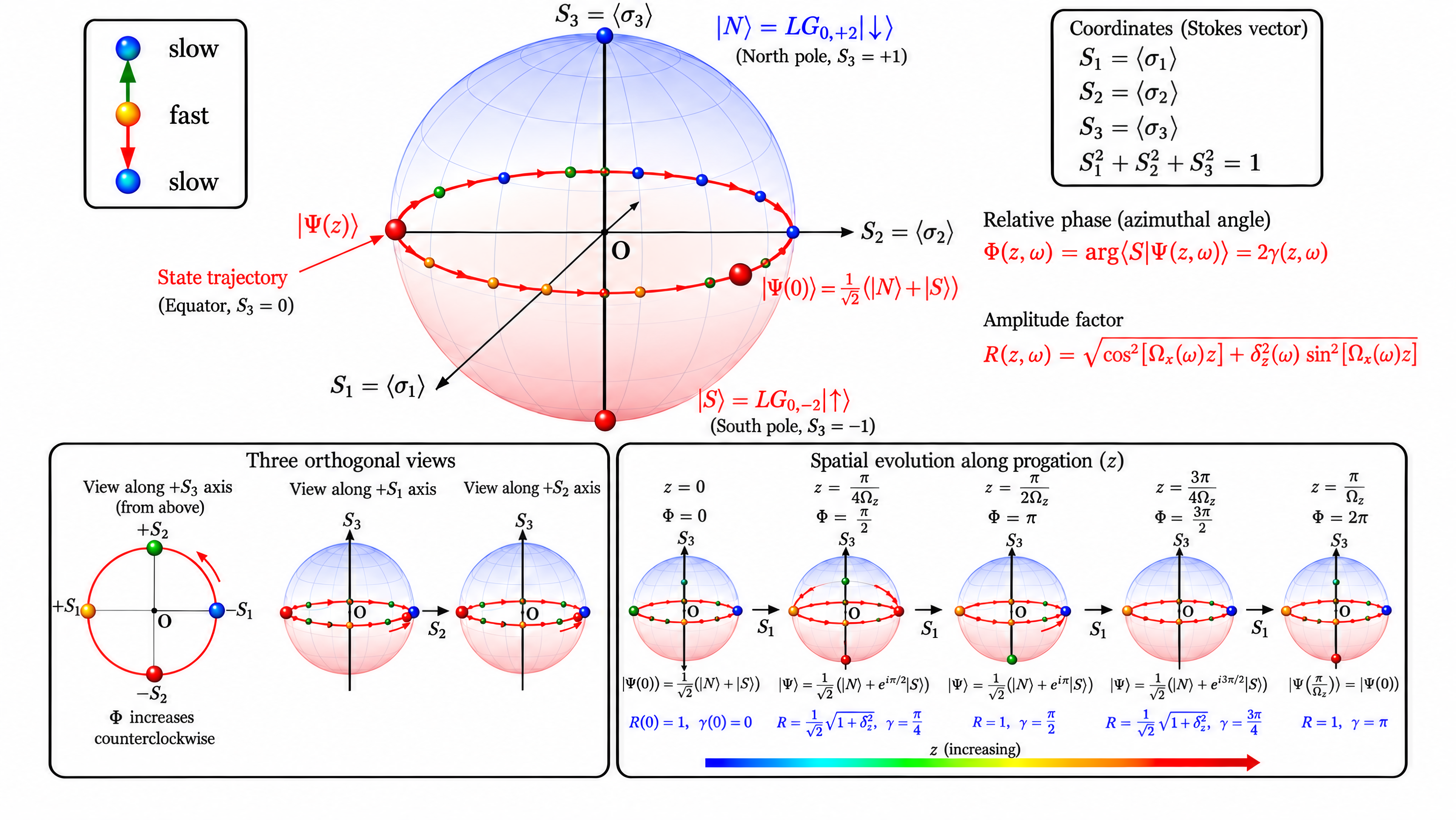}
    \caption{
    Higher-order Poincaré-sphere (HOPS) representation of the spatial evolution of the spin-orbit-coupled vector beam in the intermediate-detuning regime $0<|\delta_d|<1$, illustrated here for $\delta_d=0.3$. The north and south poles correspond to the spin-orbit basis states $\ket{N}=\ket{0,+2}\otimes\ket{\downarrow}$ and $\ket{S}=\ket{0,-2}\otimes\ket{\uparrow}$, respectively. During propagation along $z$, the state remains on the equator, $S_3=0$, while its azimuthal angle $\Phi(z)=2\gamma(z)$ evolves nonuniformly. The color-coded markers denote equal propagation-distance intervals over one spatial oscillation period $L_z=\pi/\Omega_z$. The evolution is slow near the $S_1$ axis and fast near the $S_2$ axis. The lower panels show three orthogonal projections and representative states during one complete spatial cycle.
    }
    \label{fig:HOPS}
\end{figure*}

The spatial spin-orbit interaction drives the reversible OAM annihilation transition along $z$:
\begin{equation}\label{eq:spin-orbit-transition+}
\ket{0, -2}\otimes\ket{\uparrow} \quad\rightleftharpoons\quad \ket{\circ, 0}\otimes\ket{\downarrow},
\end{equation}
where the converted zero-OAM state is a superposed radial Gaussian mode:
\begin{equation}
\ket{\circ, 0} = \frac{1}{\sqrt{6}}\left(\ket{0, 0} + 2\ket{1, 0} + \ket{2, 0}\right).
\end{equation}
Similarly, the conjugate component undergoes the complementary spatial conversion:
\begin{equation}\label{eq:spin-orbit-transition-}
\ket{0, +2}\otimes\ket{\downarrow} \quad\rightleftharpoons\quad \ket{\circ, 0}\otimes\ket{\uparrow}.
\end{equation}
Using Eq.~(\ref{eq:mateix-elements}), the corresponding spatial transition matrix elements read
\begin{equation}
\Omega_\pm = \sqrt{6}\frac{\beta}{2k_z^0}k_w^2 e^{\pm i2\kappa z} \equiv \Omega_0 e^{\pm i2\kappa z}.
\end{equation}
The $z$-dependent phase factor originates from differential spatial wavefront curvature between opposite helicities.

For the spatial conversion channel in Eq.~(\ref{eq:spin-orbit-transition+}), solving Eq.~(\ref{eq:Rabi-osc-eq}) yields the spatial mode evolution:
\begin{equation}
\begin{aligned}
{U}_+(\mathbf{r}_\perp, z) &= e^{i\Delta_\chi z} e^{-i\kappa z} A_+(z)\ket{0, -2}, \\
{U}_-(\mathbf{r}_\perp, z) &= -ie^{-i\Delta_\chi z} e^{i\kappa z} \frac{\Omega_0}{\Omega_z}\sin\left[\Omega_z z\right]\ket{\circ, 0},
\end{aligned}
\end{equation}
where $A_+(z) = \cos\left[\Omega_z z\right] - i\delta_d\sin\left[\Omega_z z\right]$, the dimensionless spatial detuning is $\delta_d = \Delta_\chi/\Omega_z$, and the effective spatial Rabi conversion frequency is
\begin{equation}
\Omega_z = \sqrt{\Delta_\chi^2 + \Omega_0^2}.
\end{equation}

The fixed radial superposition introduces a further limitation on this two-mode approximation. Conservation of $J_z$ alone does not close the radial-mode subspace. Different radial components accumulate different Gouy phases, changing their mode overlap without destroying coherence in a deterministic lossless medium. To isolate this effect, consider free propagation of the converted combination from a common waist plane with a common Rayleigh length $z_R$. After removing common transverse scaling and wavefront curvature, its coefficients acquire phases $e^{-i(2p+1)\psi}$, where $\psi=\arctan(z/z_R)$. Its overlap with the fixed coefficient combination is
\begin{equation}\label{eq:Gouy-fidelity}
F_G(z)=\left|\frac{e^{-i\psi}+4e^{-3i\psi}+e^{-5i\psi}}{6}\right|^2=\left[\frac{2+\cos(2\psi)}{3}\right]^2.
\end{equation}
Near the waist, $F_G=1-\frac43(z/z_R)^2+O[(z/z_R)^4]$. Thus $z\lesssim0.20z_R$ gives a 95\% Gouy-phase mode-overlap criterion, while $z\lesssim0.29z_R$ gives a 90\% criterion. These are conditional estimates for the fixed radial combination, not a universal coherence length or an error bound for the continuously coupled dynamics. The first conversion maximum, $L_{\mathrm{conv}}=\pi/(2\Omega_z)$, or the population period, $L_{\mathrm{osc}}=\pi/\Omega_z$, should be compared with the chosen tolerance length. Additional radial-mode leakage must independently remain small. For distinct helicity Rayleigh lengths, a conservative preliminary check uses the shorter length, but relative curvature and waist mismatch must also be evaluated. Beyond this regime, individual radial modes and their propagation-dependent couplings should be retained, or the full transverse Fourier-space evolution should be computed. Gouy phases already included in propagating LG basis functions must not be counted twice; their effects enter through the consistent projected dynamics. Mode-resolved oscillation visibility can change through radial redistribution even though the full field remains coherent.

\begin{figure*}[htpb]
    \centering
    \includegraphics[width=0.95\linewidth]{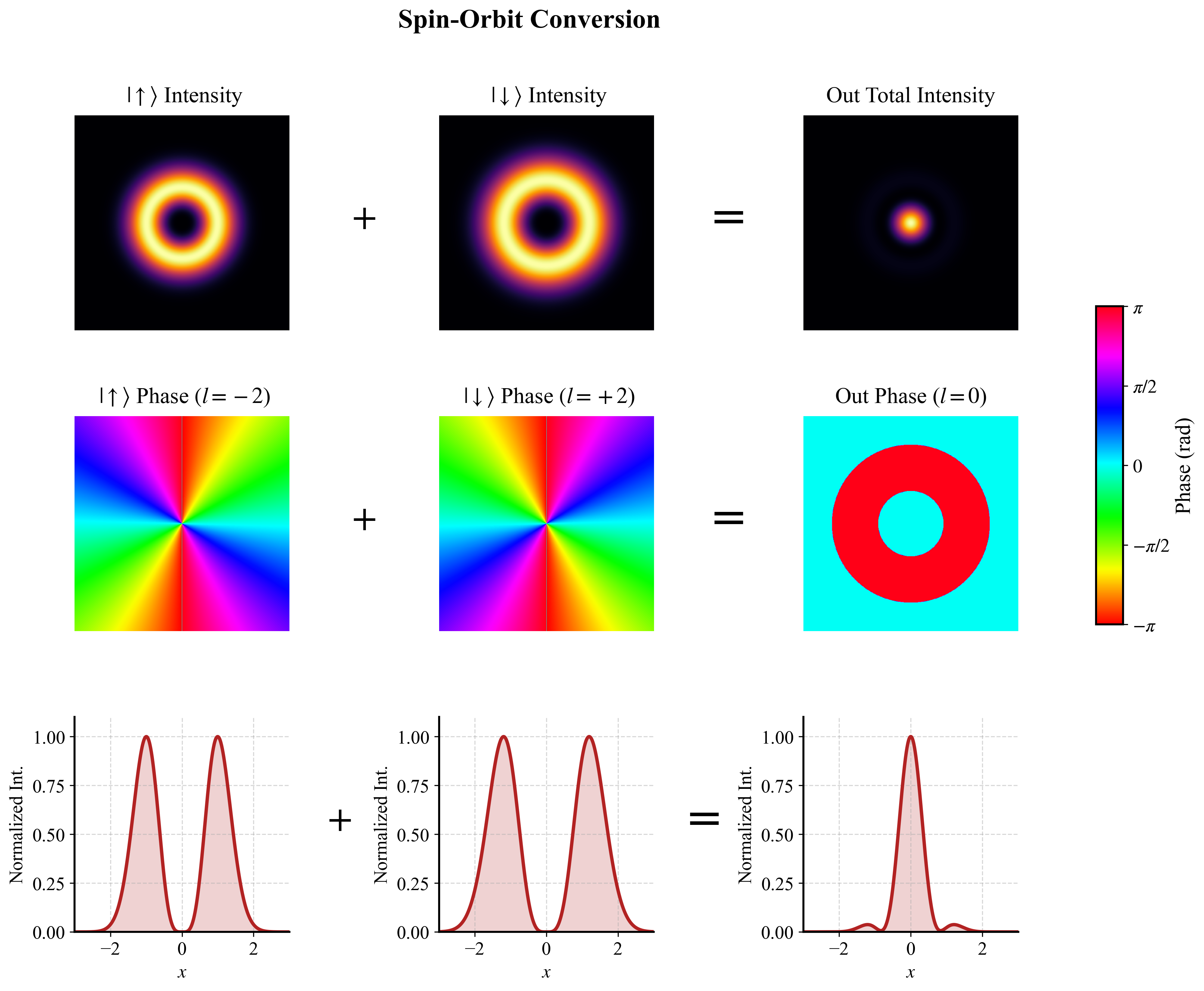}
    \caption{Intensity and phase distributions demonstrating spatial OAM annihilation in a vector vortex beam. The figure is arranged in a three-by-three grid showing normalized spatial intensity (top row), phase profiles (middle row), and corresponding 1D normalized intensity cross-sections at $y=0$ (bottom row). Formatted as a visual equation, the left and middle columns illustrate the superposed input spin-up ($\ket{\uparrow}, \ell=-2$) and spin-down ($\ket{\downarrow}, \ell=+2$) components, exhibiting doughnut-shaped intensity profiles and spiral phase fronts. The right column depicts the resulting output state after spatial spin-orbit conversion along $z$, where topological charge is completely annihilated ($\ell=0$). The 1D cross-sections clearly demonstrate this transition: the input states show a zero-intensity dark core, which collapses in the output state into an intense bright central spot (formed by a superposition of radial modes $p=0,1,2$), accompanied by chiral envelope swapping. Intensity plots use an inferno colormap ranging from zero (black) to maximum (yellow); phase plots cyclically map shifts from $-\pi$ to $\pi$.}
    \label{out1}
\end{figure*}

During spatial propagation through the dispersive chiral metamaterial, spin-orbit interaction imprints an azimuthally dependent geometric phase that completely annihilates the initial topological charge ($\ell=0$). Consequently, the topologically protected central dark core collapses, rapidly redirecting electromagnetic energy toward the optical axis into a coherent superposition of higher-order radial modes ($p=0,1,2$). Macroscopically, the converted Gaussian-like component oscillates in spatial intensity as $\delta_\beta^2\sin^2\left[\Omega_z z\right]$, where $\delta_\beta = \Omega_0/\Omega_z$ measures the efficiency of spin-degeneracy breaking. 
Simultaneously, its transverse polarization plane rotates along $z$ at rate $d\phi_p/dz = \kappa$. 

For the remaining unconverted twisted optical component with OAM $\ell=\pm 2$, parameterizing the complex amplitudes as $A_\pm(z) = R(z)e^{\mp i\gamma(z)}$ yields
\begin{equation}
\begin{aligned}
R(z) &= \sqrt{\cos^2\left[\Omega_z z\right] + \delta_d^2\sin^2\left[\Omega_z z\right]}, \\
\gamma(z) &= \tan^{-1}\left\{\delta_d\tan\left[\Omega_z z\right]\right\}.
\end{aligned}
\end{equation}
Because the LG modes carry azimuthal phase $e^{\mp i2\phi}$, the local transverse polarization orientation along $z$ evolves as
\begin{equation}
\theta(\phi, z) = 2\phi + \gamma(z) + \phi_p(z).
\end{equation}
The instantaneous spatial rotation rate of the polarization texture along the propagation axis is
\begin{equation}
\frac{d\theta}{dz} = \frac{d\gamma}{dz} = \frac{\delta_d\Omega_z}{\cos^2\left[\Omega_z z\right] + \delta_d^2\sin^2\left[\Omega_z z\right]},
\end{equation}
demonstrating that intermediate spatial detuning ($0<|\delta_d|<1$) transforms uniform polarization rotation into strongly nonuniform spatial rotation along $z$, as illustrated in Fig.~\ref{fig:R-Gamma-varing} and on the HOPS in Fig.~\ref{fig:HOPS}.

Conversely, when a linearly polarized fundamental Gaussian beam is launched into the medium:
\begin{equation}
\ket{\mathrm{in}} = \frac{1}{\sqrt{2}}\ket{0, 0}\otimes\left(e^{-i\phi_p(z)}\ket{\uparrow} + e^{i\phi_p(z)}\ket{\downarrow}\right),
\end{equation}
spatial spin-orbit coupling drives the inverse generation transitions:
\begin{equation}\label{converted}
\ket{0, 0}\otimes\ket{\uparrow} ~\rightarrow~\ket{0, +2}\otimes\ket{\downarrow}, \quad \ket{0, 0}\otimes\ket{\downarrow} ~\rightarrow~\ket{0, -2}\otimes\ket{\uparrow}.
\end{equation}
Here, the Gaussian components are periodically converted into second-order optical vortices carrying $\ell=\pm 2$ along the propagation axis $z$, with conversion intensity oscillating as $\sin^2\left[\Omega_z z\right]$, perfectly confirming the reversible nature of spatial photonic Rabi oscillations.

To validate the theoretical formalism, we numerically simulate the incident states and their evolution during spin-orbit interaction. As illustrated in Fig.~\ref{out1}, the incident field is a spin-orbit-locked vector vortex beam which is represented by a structured linear polarization state \cite{forbes2021structured}. The spin-up and spin-down components carry opposite orbital angular momenta ($\ell = -2$ and $\ell = +2$, respectively), producing characteristic spiral phase distributions. This central phase singularity enforces destructive interference, resulting in a topologically protected hollow dark core and a doughnut-shaped intensity profile for both input spin components. The spin-orbit conversion in this process is induced by the SOI in the chiral media, leading to a redistribution of the optical field and corresponding change in the LG modes, as described by Eq.(\ref{eq:spin-orbit-transition+}) and Eq. (\ref{eq:spin-orbit-transition-}).

During propagation, the chiral medium imprints an azimuthally dependent geometric phase onto the wavefront, driving a robust spin-orbit conversion  that entirely annihilates the topological charge ($\ell = 0$) \cite{PhysRevLett.96.163905}. Conversely, as described by Eq.(\ref{converted}), this spin-orbit interaction induces a dramatic topological transition: the central dark core collapses, and energy of the beam converges to the optical axis, creating an intense bright center. Because the underlying spin-orbit coupling operator is not mathematically closed for pure radial modes, this OAM annihilation inherently excites a superposition of higher-order radial modes ($p = 0, 1, 2$), which manifests as a fragmented multi-ring intensity profile in the output state \cite{shen2019optical}. 

Furthermore, the chiral medium modulates the wave numbers of the left- and right-handed states differently, causing the two input spin components to possess distinct initial beam waists. Driven by the chiral cross-coupling, the newly generated, spin-flipped outgoing states are forcefully bound to these interchanged spatial envelopes. Macroscopically, this intrinsic asymmetry manifests as a striking chiral envelope swapping that accompanies the topological charge annihilation.

Although the spin-orbit conversion locally distorts the optical wavefront and generates higher-order radial modal components, the total angular momentum of the beam remains strictly conserved throughout the entire process \cite{bliokh2015spin}. This robust interplay positions chiral media as efficient platforms for converting between photon spin and orbital degrees of freedom.

The emergence of these spatial Rabi dynamics is strictly conditional. Breaking the spin-degenerate condition provides the fundamental transverse spatial torque required to flip the photon spin and transfer the angular momentum to the spatial wavefront. The condition for observing spatial spin-orbit conversion is governed by the frequency-dependent parameter $\beta$. The spin-degenerate regime ($\beta=0$) strictly requires matched transverse-to-longitudinal impedance anisotropy:
\begin{equation}\label{eq:matched_anisotropy}
\frac{\epsilon_t}{\epsilon_z} = \frac{\mu_t}{\mu_z}.
\end{equation}
Because Eq.~(\ref{eq:matched_anisotropy}) is significantly weaker than complete isotropy  $\boldsymbol{\epsilon}=\boldsymbol{\mu}$, spatial Rabi oscillations can be readily engineered in artificial dielectric metasurfaces or twisted bianisotropic microstructures \cite{gorlach2018far}.

It is crucial to emphasize that the emergence of this spatial Rabi dynamics highlights the fundamental physical difference between our exact optical Dirac framework and conventional phenomenological models. In previous studies, photonic Rabi oscillations are commonly described using an effective pseudospin Hamiltonian driven by an externally engineered synthetic magnetic field. Here, we reveal its exact microscopic electromagnetic origin: the effective driving field emerges intrinsically from the $\gamma_5$-extended optical Dirac Hamiltonian. The parameter $\beta$, which captures the electric-magnetic mismatch, fundamentally breaks the electromagnetic duality and acts as the generator for this intrinsic helicity-orbital coupling. 

Therefore, Eq.~(\ref{eq:matched_anisotropy}) is not just a mathematical singularity, but a highly specific, experimentally accessible prediction. Within the effective model, frequency tuning that breaks or satisfies this matching condition can enable or suppress the helicity-mixing term. Observable oscillations additionally require sufficient coupling relative to detuning and loss, together with the propagation-validity conditions stated above. This provides a clear experimental roadmap for manipulating angular momentum conversion without relying on complex, external periodic modulations.

In a spin-degenerate medium, no conversion between SAM and OAM occurs. Nevertheless, transitions satisfying the selection rule $\Delta \ell = \pm 2$ may arise due to quadrupolar interactions between structured light and the medium. A notable example is the propagation of twisted light in a gravitational-wave background \cite{Wu_2022}, which can be regarded as a spin-degenerate medium characterized by equal inverse permittivity and permeability tensors, $\boldsymbol{\epsilon}^{-1} = \boldsymbol{\mu}^{-1} = \boldsymbol{\gamma}$, where $\boldsymbol{\gamma} = \{g_{ij}\}$ represents the effective spatial metric. In this setting, $\Delta \ell = \pm 2$ transitions emerge from quadrupole couplings of the form $Q_{\mp}$ $\hat{k}_{\pm}^2$, where $Q_{\pm}$ denote the tensor excitation modes associated with the gravitational wave and are directly linked to helicity-2 gravitational perturbations. These processes reflect the coupling between the spin-2 nature of gravitational waves and the orbital structure of twisted light. If such angular-momentum transfer between gravitons and photons could be experimentally detected through future high-precision optical measurements, it would provide a direct probe of the quantum nature of gravity.      

From the viewpoint of material design, the effective two-level Hamiltonian establishes a direct mapping between the macroscopic constitutive parameters and the resulting helicity–orbital dynamics. The chirality parameters \(\chi_{t,z}\) primarily control the helicity-dependent propagation constants and the associated detuning, whereas the electric–magnetic anisotropy mismatch,
\(
{\epsilon_t}/{\epsilon_z}-{\mu_t}/{\mu_z},
\) determines the transverse helicity-mixing interaction through the coupling parameter \(\beta\). Efficient spin–orbit conversion therefore requires a deliberately non-spin-degenerate medium with \(\beta\neq0\), together with a sufficiently small effective detuning. For example, a circularly polarized Gaussian input state \(|0,0\rangle\otimes|\uparrow\rangle\) can be coherently coupled to the vortex state \(|0,2\rangle\otimes|\downarrow\rangle\), with the conversion length determined by the corresponding Rabi frequency. The desired oscillation strength and conversion length can thus be engineered by tailoring the electric, magnetic, and magnetoelectric responses of the medium.
Such constitutive responses can, in practice, be implemented using dielectric metasurfaces or artificial bianisotropic microstructures with appropriately engineered anisotropy and spatially twisted configurations \cite{gorlach2018far,zhang2024metasurface}. These platforms provide a possible route for independently tailoring the electric–magnetic mismatch and chiral response, and hence for controlling the effective transverse coupling and helicity detuning predicted by our theory. Experimentally, the resulting helicity–orbital conversion can be characterized through full-Stokes polarimetry combined with coherent interferometric imaging, which provides access to both the polarization evolution and the phase structure of the generated OAM modes. In regimes where the conversion is accompanied by the topological transition discussed above, the same measurements can also reveal the evolution and collapse of the corresponding optical singularities. More broadly, this constitutive-parameter-level design principle provides a route toward controllable SAM–OAM conversion and the manipulation of structured-light topological states, with potential applications to high-dimensional quantum-state generation \cite{liang2025metasurface,shi2021quantum} and all-optical angular-momentum routing \cite{alarcon2023all,qiu2020efficient,luo2017synthetic}.

\subsection{Material Dissipation and Observable Propagation Range}\label{sec:loss-validity}

The lossless model isolates the coherent conversion mechanism, whereas experimental observability depends on the competition between coupling and attenuation. The Kramers--Kronig relations connect dispersion and absorption across the spectrum, but do not require strong absorption at every frequency with appreciable dispersion. Resonant enhancement should therefore be assessed through the coupling-to-loss ratio and effective detuning, rather than the coupling strength alone.

For a leading-order estimate, we retain the two-mode basis and model absorption by diagonal attenuation. After removing a common real propagation phase and choosing the relative modal phase so that the coherent coupling $g$ is real and nonnegative, the amplitudes obey
\begin{equation}\label{eq:lossy-two-mode}
i\frac{d}{dz}\binom{C_+}{C_-}=
\begin{pmatrix}
\Delta-i\gamma_+/2 & g\\
g &-\Delta-i\gamma_-/2
\end{pmatrix}\binom{C_+}{C_-}.
\end{equation}
Here $\Delta$ is the real effective detuning, and $g=|\Omega_0|$ in the constant-coupling limit of Sec.~4.2. The amplitudes are normalized so that $P_\pm=|C_\pm|^2$ represent modal powers. All coefficients are taken to be independent of $z$, with $\gamma_\pm\geq0$ denoting power attenuation coefficients: an uncoupled channel has $P_\pm(z)=P_\pm(0)e^{-\gamma_\pm z}$. For a physical modal field proportional to $e^{ik_\pm z}$, $\gamma_\pm=2\operatorname{Im}k_\pm$. Equation~(\ref{eq:lossy-two-mode}) gives $d(P_++P_-)/dz=-\gamma_+P_+-\gamma_-P_-$, consistent with passive attenuation. It neglects dissipative cross-coupling and loss-induced changes of the mode profiles; these require a projection using the complex constitutive tensors when they are appreciable.

For equal attenuation, $\gamma_+=\gamma_-=\gamma$, and initial excitation of the $+$ channel, the converted power is
\begin{equation}\label{eq:lossy-conversion}
\frac{P_-(z)}{P_{\mathrm{in}}}
=e^{-\gamma z}\frac{g^2}{\Omega^2}\sin^2(\Omega z),\quad
\Omega=\sqrt{\Delta^2+g^2}.
\end{equation}
The total power decays as $P_++P_-=P_{\mathrm{in}}e^{-\gamma z}$, while the conversion fraction $P_-/(P_++P_-)$ retains its lossless form. Thus common absorption damps the absolute signal at the power decay rate $\gamma$ without intrinsically changing the normalized conversion curve. Experimental visibility nevertheless deteriorates as the signal approaches the noise floor. The maximum normalized conversion is $g^2/(\Delta^2+g^2)$, so a short oscillation period alone does not ensure efficient transfer.

Define the power attenuation length $L_{\mathrm{loss}}=1/\gamma$ and population oscillation period $L_{\mathrm{osc}}=\pi/\Omega$. Weak damping over one cycle requires $L_{\mathrm{osc}}\ll L_{\mathrm{loss}}$; over $N$ cycles it requires $NL_{\mathrm{osc}}\ll L_{\mathrm{loss}}$. For example, retaining at least 90\% of the power after one cycle requires $L_{\mathrm{loss}}\geq L_{\mathrm{osc}}/[-\ln(0.9)]\simeq9.5L_{\mathrm{osc}}$. If $T_{\min}$ is a minimum detectable total-power fraction, attenuation alone gives the upper range $L_{\max}=L_{\mathrm{loss}}\ln(1/T_{\min})$. Detecting the converted channel additionally depends on its conversion fraction and the measurement noise; this total-power criterion is therefore not sufficient by itself.

Differential attenuation also modifies the normalized dynamics. With $\bar\gamma=(\gamma_++\gamma_-)/2$ and $\delta\gamma=\gamma_+-\gamma_-$, the eigenvalues of the effective propagation generator are $-i\bar\gamma/2\pm\widetilde\Omega$, where $\widetilde\Omega=\sqrt{g^2+(\Delta-i\delta\gamma/4)^2}$. At zero detuning, the oscillatory regime requires $g>|\delta\gamma|/4$, and its population period is $\pi/\sqrt{g^2-(\delta\gamma/4)^2}$. At and below this threshold, the sinusoidal exchange is replaced by critically damped or nonoscillatory transfer within this model. Unequal attenuation biases the surviving populations, while the observable number of cycles remains limited by the transmitted signal. For nonzero detuning, $\widetilde\Omega$ is generally complex and the simple resonant threshold does not apply.

An operating point on a resonance flank or within a low-loss window may therefore be preferable to the point of maximum material response. Quantitative attenuation lengths require measured or calculated complex constitutive parameters; the lossless model alone provides no material-specific observability estimate. Physical dissipation is distinct from the representation-induced non-Hermiticity of the conservative formulation. The usable propagation range must also satisfy the radial-mode restrictions of Sec.~\ref{rabi} and the paraxial conditions in ~\ref{sec:AppB}.

\section{Conclusion}\label{sec5}

In this study, we generalize the optical Dirac theory \cite{PhysRevA.106.043513} to structured light propagating in chiral optical media. The central novelty of our formulation is a helicity-space reformulation of Maxwell’s equations based on a complex electromagnetic field vector distinct from the conventional Riemann-Silberstein representation. This construction naturally organizes the electromagnetic field into a Dirac-like spinor structure and yields a non-Hermitian chiral $\gamma_5$-extension of the optical Dirac equation, formally analogous to the dynamics of massive fermions with anomalous magnetic moments in an external pseudomagnetic field. In this way, the intrinsic spin-orbit dynamics of light are rendered fully explicit directly at the level of Maxwell's equations.

Beyond the spin-degenerate limit, the optical Dirac equation admits four eigenvalues and the corresponding eigenstates, representing forward- and backward-propagating circularly polarized modes with opposite helicities. We analyze how this spectrum depends on the constitutive parameters of the chiral medium, with particular emphasis on the role of chirality in controlling helicity splitting in the resulting non-Hermitian optical Hamiltonian.

To describe structured light carrying orbital angular momentum, we further project the optical Dirac equation onto the biorthonormal basis formed by the two positive-energy branches. This projection yields an effective two-component Schr\"odinger-like equation for light propagating in chiral photonic media. The resulting theory shows that spin-orbit coupling drives coherent conversion between SAM and OAM during propagation. In particular, a spin-up to spin-down transition transfers two units of SAM into OAM, whereas the inverse transition converts two units of OAM back into SAM. Crucially, because this coupling stems fundamentally from the breaking of the strict spin-degeneracy condition, where off-diagonal orbital ladder operators emerge, the SAM-to-OAM conversion occurs while strictly conserving the total angular momentum of the optical field.

An exact solution of this two-level system in wave-vector space establishes a robust photonic counterpart of conventional atomic Rabi oscillations \cite{rabiPhysRev.51.652,rabiobservationPhysRevLett.58.353,shandarova2009experimental,liu2023spin,liu2023higher}. In this framework, the two photon helicity states, coupled through the spin-orbit interaction of light, play roles analogous to the ground and excited states of an atomic system. Most notably, the present system exhibits a distinctive form of Rabi-like oscillation characterized by an inverted driving mechanism: instead of an optical field driving transitions between material states, the chiral medium acts as an effective background field that drives coherent oscillations between distinct photonic spin and orbital angular-momentum states.

These dynamics, governed by quadrupole interactions within the effective two-component optical theory, enable efficient and reversible angular-momentum conversion. In particular, incident vector beams whose two spin components carry opposite topological charges can be coherently transformed into Gaussian-like modes with vanishing OAM, and vice versa. At the same time, the evolution of the polarization texture is controlled by a chirality-induced detuning parameter, giving rise to dynamical regimes ranging from purely periodic intensity modulation to uniform or nonuniform polarization rotation, while preserving the underlying second-order topological structure of the vector beam \cite{zhan2009cylindrical}. These results demonstrate that chiral media provide an alternative platform for realizing Rabi dynamics and open new routes for the spatial control of structured vector optical fields and spin-associated topological charges.

Ultimately, the optical Dirac equation provides a compact and systematic framework for analyzing light propagation in chiral photonic media. The theory not only clarifies the electromagnetic origin of spin-orbit coupling in such systems, but also offers a useful route for designing chiral photonic structures for angular-momentum control, polarization engineering, and helicity-sensitive optical manipulation\cite{PhysRevB.88.085111,khorasaninejad2017metalenses,maguid2016photonic}.

\Acknowledgements{F.L.L. is supported by the National Key R\&D Program of China through Grant No.~2020YFC2201400.}

\InterestConflict{The authors declare that they have no conflict of interest.}

\bibliographystyle{scpma} 
\bibliography{Photonic_Rabi_chiral_media}

@article{PhysRevA.106.043513,
  title = {Four-vector optical {Dirac} equation and spin-orbit interaction of structured light},
  author = {Feng, Longlong and Wu, Qianfan},
  journal = {\href{https://doi.org/10.1103/PhysRevA.106.043513}{Phys. Rev. A}},
  volume = {106},
  number = {4},
  pages = {043513},
  year = {2022},
  month = {Oct},
  publisher = {American Physical Society},
  doi = {10.1103/PhysRevA.106.043513}
}

@article{khanikaev2013photonic,
  title = {Photonic topological insulators},
  author = {Khanikaev, Alexander B. and Hossein Mousavi, S. and Tse, Wang-Kong and Kargarian, Mehdi and MacDonald, Allan H. and Shvets, Gennady},
  journal = {\href{https://doi.org/10.1038/nmat3520}{Nat. Mater.}},
  volume = {12},
  number = {3},
  pages = {233--239},
  year = {2013},
  publisher = {Nature Publishing Group},
  doi = {10.1038/nmat3520}
}

@article{PhysRevLett.102.023901,
  title = {Negative Refractive Index in Chiral Metamaterials},
  author = {Zhang, Shuang and Park, Yong-Shik and Li, Jensen and Lu, Xinchao and Zhang, Weili and Zhang, Xiang},
  journal = {\href{https://doi.org/10.1103/PhysRevLett.102.023901}{Phys. Rev. Lett.}},
  volume = {102},
  number = {2},
  pages = {023901},
  year = {2009},
  month = {Jan},
  publisher = {American Physical Society},
  doi = {10.1103/PhysRevLett.102.023901}
}

@article{chern2023photonic,
  title = {Photonic helicoid-like surface states in chiral metamaterials},
  author = {Chern, Ruey-Lin},
  journal = {\href{https://doi.org/10.1038/s41598-023-40926-8}{Sci. Rep.}},
  volume = {13},
  number = {1},
  pages = {13934},
  year = {2023},
  publisher = {Nature Publishing Group},
  doi = {10.1038/s41598-023-40926-8}
}

@article{Wu_2022,
  title = {Testing the Wave-Particle Duality of Gravitational Wave Using the Spin-Orbital-{Hall} Effect of Structured Light},
  author = {Wu, Qianfan and Zhu, Weishan and Feng, Longlong},
  journal = {\href{https://doi.org/10.3390/universe8100535}{Universe}},
  volume = {8},
  number = {10},
  pages = {535},
  year = {2022},
  month = {Oct},
  publisher = {MDPI AG},
  doi = {10.3390/universe8100535}
}

@article{PhysRevLett.117.057401,
  title = {Hyperbolic {Weyl} Point in Reciprocal Chiral Metamaterials},
  author = {Xiao, Meng and Lin, Qian and Fan, Shanhui},
  journal = {\href{https://doi.org/10.1103/PhysRevLett.117.057401}{Phys. Rev. Lett.}},
  volume = {117},
  number = {5},
  pages = {057401},
  year = {2016},
  month = {Jul},
  publisher = {American Physical Society},
  doi = {10.1103/PhysRevLett.117.057401}
}

@article{ma2022topological,
  title = {Topological photonics in metamaterials},
  author = {Ma, Shaojie and Yang, Biao and Zhang, Shuang},
  journal = {\href{https://doi.org/10.3788/PI.2022.R02}{Photonics Insights}},
  volume = {1},
  number = {1},
  pages = {R02},
  year = {2022},
  publisher = {Society of Photo-Optical Instrumentation Engineers},
  doi = {10.3788/PI.2022.R02}
}

@article{Rubinsztein-Dunlop2016,
  title = {Roadmap on structured light},
  author = {Rubinsztein-Dunlop, H. and Forbes, A. and Berry, M. V. and Dennis, M. R. and Andrews, D. L. and Mansuripur, M. and others},
  journal = {\href{https://doi.org/10.1088/2040-8978/19/1/013001}{J. Opt.}},
  volume = {19},
  number = {1},
  pages = {013001},
  year = {2017},
  publisher = {IOP Publishing},
  doi = {10.1088/2040-8978/19/1/013001}
}

@article{Angelsky2020,
  title = {Structured Light: Ideas and Concepts},
  author = {Angelsky, Oleg V. and Bekshaev, Aleksandr Y. and Hanson, Steen G. and Zenkova, Claudia Yu. and Mokhun, Igor I. and Zheng, Jun},
  journal = {\href{https://doi.org/10.3389/fphy.2020.00114}{Front. Phys.}},
  volume = {8},
  pages = {114},
  year = {2020},
  publisher = {Frontiers Media SA},
  doi = {10.3389/fphy.2020.00114}
}

@misc{andrews2013angular,
  author = {Andrews, D. L. and Babiker, M.},
  title = {The Angular Momentum of Light},
  note = {Cambridge: Cambridge University Press, 2013},
  doi = {10.1017/CBO9780511795213}
}

@article{bliokh2015spin,
  title = {Spin--orbit interactions of light},
  author = {Bliokh, Konstantin Yu. and Rodr{\'\i}guez-Fortu{\~n}o, Francisco J. and Nori, Franco and Zayats, Anatoly V.},
  journal = {\href{https://doi.org/10.1038/nphoton.2015.201}{Nat. Photonics}},
  volume = {9},
  number = {12},
  pages = {796--808},
  year = {2015},
  publisher = {Nature Publishing Group},
  doi = {10.1038/nphoton.2015.201}
}

@article{wang2018magnetic,
  title = {Magnetic spin--orbit interaction of light},
  author = {Wang, Mengjia and Zhang, Hongyi and Kovalevich, Tatiana and Salut, Roland and Kim, Myun-Sik and Suarez, Miguel Angel and Bernal, Maria-Pilar and Herzig, Hans-Peter and Lu, Huihui and Grosjean, Thierry},
  journal = {\href{https://doi.org/10.1038/s41377-018-0018-9}{Light Sci. Appl.}},
  volume = {7},
  number = {1},
  pages = {24},
  year = {2018},
  publisher = {Nature Publishing Group},
  doi = {10.1038/s41377-018-0018-9}
}

@article{shao2018spin,
  title = {Spin-orbit interaction of light induced by transverse spin angular momentum engineering},
  author = {Shao, Zengkai and Zhu, Jiangbo and Chen, Yujie and Zhang, Yanfeng and Yu, Siyuan},
  journal = {\href{https://doi.org/10.1038/s41467-018-03237-5}{Nat. Commun.}},
  volume = {9},
  number = {1},
  pages = {926},
  year = {2018},
  publisher = {Nature Publishing Group},
  doi = {10.1038/s41467-018-03237-5}
}

@article{PhysRevD.90.105018,
  title = {Relativistic Particle and Relativistic Fluids: Magnetic Moment and Spin-Orbit Interactions},
  author = {Karabali, Dimitra and Nair, V. P.},
  journal = {\href{https://doi.org/10.1103/PhysRevD.90.105018}{Phys. Rev. D}},
  volume = {90},
  number = {10},
  pages = {105018},
  year = {2014},
  month = {Nov},
  publisher = {American Physical Society},
  doi = {10.1103/PhysRevD.90.105018}
}

@article{Spavieri_2015,
  title = {Origin of the spin--orbit interaction},
  author = {Spavieri, Gianfranco and Mansuripur, Masud},
  journal = {\href{https://doi.org/10.1088/0031-8949/90/8/085501}{Phys. Scr.}},
  volume = {90},
  number = {8},
  pages = {085501},
  year = {2015},
  month = {Jun},
  publisher = {IOP Publishing},
  doi = {10.1088/0031-8949/90/8/085501}
}

@article{PhysRevC.94.024304,
  title = {Spin-orbit interaction in relativistic nuclear structure models},
  author = {Ebran, J.-P. and Mutschler, A. and Khan, E. and Vretenar, D.},
  journal = {\href{https://doi.org/10.1103/PhysRevC.94.024304}{Phys. Rev. C}},
  volume = {94},
  number = {2},
  pages = {024304},
  year = {2016},
  month = {Aug},
  publisher = {American Physical Society},
  doi = {10.1103/PhysRevC.94.024304}
}

@article{PhysRevA.97.043840,
  title = {Relativistic spin-orbit interactions of photons and electrons},
  author = {Smirnova, D. A. and Travin, V. M. and Bliokh, K. Y. and Nori, F.},
  journal = {\href{https://doi.org/10.1103/PhysRevA.97.043840}{Phys. Rev. A}},
  volume = {97},
  number = {4},
  pages = {043840},
  year = {2018},
  month = {Apr},
  publisher = {American Physical Society},
  doi = {10.1103/PhysRevA.97.043840}
}

@article{PhysRevLett.106.106602,
  title = {Spin-Orbit Coupling, Spin Relaxation, and Spin Diffusion in Organic Solids},
  author = {Yu, Z. G.},
  journal = {\href{https://doi.org/10.1103/PhysRevLett.106.106602}{Phys. Rev. Lett.}},
  volume = {106},
  number = {10},
  pages = {106602},
  year = {2011},
  month = {Mar},
  publisher = {American Physical Society},
  doi = {10.1103/PhysRevLett.106.106602}
}

@article{PhysRevB.85.115201,
  title = {Spin-orbit coupling and its effects in organic solids},
  author = {Yu, Z. G.},
  journal = {\href{https://doi.org/10.1103/PhysRevB.85.115201}{Phys. Rev. B}},
  volume = {85},
  number = {11},
  pages = {115201},
  year = {2012},
  month = {Mar},
  publisher = {American Physical Society},
  doi = {10.1103/PhysRevB.85.115201}
}

@article{li2017stripe,
  title = {A stripe phase with supersolid properties in spin--orbit-coupled {Bose--Einstein} condensates},
  author = {Li, Jun-Ru and Lee, Jeongwon and Huang, Wujie and Burchesky, Sean and Shteynas, Boris and Top, Furkan {\c{C}}a{\u{g}}r{\i} and Jamison, Alan O. and Ketterle, Wolfgang},
  journal = {\href{https://doi.org/10.1038/nature21431}{Nature}},
  volume = {543},
  number = {7643},
  pages = {91--94},
  year = {2017},
  publisher = {Nature Publishing Group},
  doi = {10.1038/nature21431}
}

@article{PhysRevLett.96.073903,
  title = {Conservation of Angular Momentum, Transverse Shift, and Spin {Hall} Effect in Reflection and Refraction of an Electromagnetic Wave Packet},
  author = {Bliokh, Konstantin Yu. and Bliokh, Yury P.},
  journal = {\href{https://doi.org/10.1103/PhysRevLett.96.073903}{Phys. Rev. Lett.}},
  volume = {96},
  number = {7},
  pages = {073903},
  year = {2006},
  month = {Feb},
  publisher = {American Physical Society},
  doi = {10.1103/PhysRevLett.96.073903}
}

@article{PhysRevLett.112.113902,
  title = {Observation of the Geometric Spin {Hall} Effect of Light},
  author = {Korger, Jan and Aiello, Andrea and Chille, Vanessa and Banzer, Peter and Wittmann, Christoffer and Lindlein, Norbert and Marquardt, Christoph and Leuchs, Gerd},
  journal = {\href{https://doi.org/10.1103/PhysRevLett.112.113902}{Phys. Rev. Lett.}},
  volume = {112},
  number = {11},
  pages = {113902},
  year = {2014},
  month = {Mar},
  publisher = {American Physical Society},
  doi = {10.1103/PhysRevLett.112.113902}
}

@article{PhysRev.50.115,
  title = {Mechanical Detection and Measurement of the Angular Momentum of Light},
  author = {Beth, Richard A.},
  journal = {\href{https://doi.org/10.1103/PhysRev.50.115}{Phys. Rev.}},
  volume = {50},
  number = {2},
  pages = {115--125},
  year = {1936},
  month = {Jul},
  publisher = {American Physical Society},
  doi = {10.1103/PhysRev.50.115}
}

@article{hosten2008observation,
  title = {Observation of the spin {Hall} effect of light via weak measurements},
  author = {Hosten, Onur and Kwiat, Paul},
  journal = {\href{https://doi.org/10.1126/science.1152697}{Science}},
  volume = {319},
  number = {5864},
  pages = {787--790},
  year = {2008},
  publisher = {American Association for the Advancement of Science},
  doi = {10.1126/science.1152697}
}

@article{PhysRevLett.102.123903,
  title = {Spin {Hall} Effect of Light in Spherical Geometry},
  author = {Haefner, D. and Sukhov, S. and Dogariu, A.},
  journal = {\href{https://doi.org/10.1103/PhysRevLett.102.123903}{Phys. Rev. Lett.}},
  volume = {102},
  number = {12},
  pages = {123903},
  year = {2009},
  month = {Mar},
  publisher = {American Physical Society},
  doi = {10.1103/PhysRevLett.102.123903}
}

@article{hermosa2011spin,
  title = {Spin {Hall} effect of light in metallic reflection},
  author = {Hermosa, N. and Nugrowati, A. M. and Aiello, Andrea and Woerdman, J. P.},
  journal = {\href{https://doi.org/10.1364/OL.36.003200}{Opt. Lett.}},
  volume = {36},
  number = {16},
  pages = {3200--3202},
  year = {2011},
  publisher = {Optica Publishing Group},
  doi = {10.1364/OL.36.003200}
}

@article{bliokh2004modified,
  title = {Modified geometrical optics of a smoothly inhomogeneous isotropic medium: The anisotropy, {Berry} phase, and the optical {Magnus} effect},
  author = {Bliokh, K. Yu. and Bliokh, Yu. P.},
  journal = {\href{https://doi.org/10.1103/PhysRevE.70.026605}{Phys. Rev. E}},
  volume = {70},
  number = {2},
  pages = {026605},
  year = {2004},
  publisher = {American Physical Society},
  doi = {10.1103/PhysRevE.70.026605}
}

@article{bliokh2008geometrodynamics,
  title = {Geometrodynamics of spinning light},
  author = {Bliokh, Konstantin Y. and Niv, Avi and Kleiner, Vladimir and Hasman, Erez},
  journal = {\href{https://doi.org/10.1038/nphoton.2008.229}{Nat. Photonics}},
  volume = {2},
  number = {12},
  pages = {748--753},
  year = {2008},
  publisher = {Nature Publishing Group},
  doi = {10.1038/nphoton.2008.229}
}

@article{luo2017broadband,
  title = {Broadband spin {Hall} effect of light in single nanoapertures},
  author = {Luo, Xian-Gang and Pu, Ming-Bo and Li, Xiong and Ma, Xiao-Liang},
  journal = {\href{https://doi.org/10.1038/lsa.2016.276}{Light Sci. Appl.}},
  volume = {6},
  number = {6},
  pages = {e16276},
  year = {2017},
  publisher = {Nature Publishing Group},
  doi = {10.1038/lsa.2016.276}
}

@article{Ling_2017,
  title = {Recent advances in the spin {Hall} effect of light},
  author = {Ling, Xiaohui and Zhou, Xinxing and Huang, Kun and Liu, Yachao and Qiu, Cheng-Wei and Luo, Hailu and Wen, Shuangchun},
  journal = {\href{https://doi.org/10.1088/1361-6633/aa5397}{Rep. Prog. Phys.}},
  volume = {80},
  number = {6},
  pages = {066401},
  year = {2017},
  month = {Mar},
  publisher = {IOP Publishing},
  doi = {10.1088/1361-6633/aa5397}
}

@article{PhysRevD.102.024075,
  title = {Gravitational spin {Hall} effect of light},
  author = {Oancea, Marius A. and Joudioux, J\'er\'emie and Dodin, I. Y. and Ruiz, D. E. and Paganini, Claudio F. and Andersson, Lars},
  journal = {\href{https://doi.org/10.1103/PhysRevD.102.024075}{Phys. Rev. D}},
  volume = {102},
  number = {2},
  pages = {024075},
  year = {2020},
  month = {Jul},
  publisher = {American Physical Society},
  doi = {10.1103/PhysRevD.102.024075}
}

@article{yu2021spin,
  title = {Spin {Hall} effect of light based on a surface plasmonic platform},
  author = {Yu, Xiantong and Wang, Xin and Li, Zhao and Zhao, Litao and Zhou, Feifan and Qu, Junle and Song, Jun},
  journal = {\href{https://doi.org/10.1515/nanoph-2021-0217}{Nanophotonics}},
  volume = {10},
  number = {12},
  pages = {3031--3048},
  year = {2021},
  publisher = {De Gruyter},
  doi = {doi:10.1515/nanoph-2021-0217}
}

@article{kim2023,
  title = {Spin {Hall} Effect of Light: From Fundamentals To Recent Advancements},
  author = {Kim, Minkyung and Yang, Younghwan and Lee, Dasol and Kim, Yeseul and Kim, Hongyoon and Rho, Junsuk},
  journal = {\href{https://doi.org/10.1002/lpor.202200046}{Laser Photonics Rev.}},
  volume = {17},
  number = {1},
  pages = {2200046},
  year = {2023},
  doi = {10.1002/lpor.202200046}
}

@article{Zhang_2014,
  title = {Orbit-orbit interaction and photonic orbital {Hall} effect in reflection of a light beam},
  author = {Zhang, J. and Zhou, X. X. and Ling, X. H. and Chen, S. Z. and Luo, H. L. and Wen, S. C.},
  journal = {\href{https://doi.org/10.1088/1674-1056/23/6/064215}{Chin. Phys. B}},
  volume = {23},
  number = {6},
  pages = {064215},
  year = {2014},
  month = {Apr},
  doi = {10.1088/1674-1056/23/6/064215}
}

@article{PhysRevLett.123.243904,
  title = {Spin-Orbit Optical {Hall} Effect},
  author = {Fu, Shenhe and Guo, Chaoheng and Liu, Guohua and Li, Yongyao and Yin, Hao and Li, Zhen and Chen, Zhenqiang},
  journal = {\href{https://doi.org/10.1103/PhysRevLett.123.243904}{Phys. Rev. Lett.}},
  volume = {123},
  number = {24},
  pages = {243904},
  year = {2019},
  month = {Dec},
  publisher = {American Physical Society},
  doi = {10.1103/PhysRevLett.123.243904}
}

@article{porfirev2023vectorial,
  title = {Vectorial spin-orbital {Hall} effect of light upon tight focusing and its experimental observation in azopolymer films},
  author = {Porfirev, Alexey and Khonina, Svetlana and Ustinov, Andrey and Ivliev, Nikolay and Golub, Ilya},
  journal = {\href{https://doi.org/10.29026/oes.2023.230014}{Opto-Electron. Sci.}},
  volume = {2},
  number = {7},
  pages = {230014},
  year = {2023},
  publisher = {Opto-Electronic Science},
  doi = {10.29026/oes.2023.230014}
}

@article{choi2023observation,
  title = {Observation of the orbital {Hall} effect in a light metal {Ti}},
  author = {Choi, Young-Gwan and Jo, Daegeun and Ko, Kyung-Hun and Go, Dongwook and Kim, Kyung-Han and Park, Hee Gyum and Kim, Changyoung and Min, Byoung-Chul and Choi, Gyung-Min and Lee, Hyun-Woo},
  journal = {\href{https://doi.org/10.1038/s41586-023-06101-9}{Nature}},
  volume = {619},
  number = {7968},
  pages = {52--56},
  year = {2023},
  publisher = {Nature Publishing Group},
  doi = {10.1038/s41586-023-06101-9}
}

@article{LUO2012864,
  title = {Spin-to-orbital angular momentum conversion in spin {Hall} effect of light},
  author = {Luo, Hailu and Wen, Shuangchun and Shu, Weixing and Fan, Dianyuan},
  journal = {\href{https://doi.org/10.1016/j.optcom.2011.10.014}{Opt. Commun.}},
  volume = {285},
  number = {6},
  pages = {864--871},
  year = {2012},
  doi = {10.1016/j.optcom.2011.10.014}
}

@article{PhysRevLett.96.163905,
  title = {Optical Spin-to-Orbital Angular Momentum Conversion in Inhomogeneous Anisotropic Media},
  author = {Marrucci, L. and Manzo, C. and Paparo, D.},
  journal = {\href{https://doi.org/10.1103/PhysRevLett.96.163905}{Phys. Rev. Lett.}},
  volume = {96},
  number = {16},
  pages = {163905},
  year = {2006},
  month = {Apr},
  publisher = {American Physical Society},
  doi = {10.1103/PhysRevLett.96.163905}
}

@article{PhysRevLett.99.073901,
  title = {Spin-to-Orbital Angular Momentum Conversion in a Strongly Focused Optical Beam},
  author = {Zhao, Yiqiong and Edgar, J. Scott and Jeffries, Gavin D. M. and McGloin, David and Chiu, Daniel T.},
  journal = {\href{https://doi.org/10.1103/PhysRevLett.99.073901}{Phys. Rev. Lett.}},
  volume = {99},
  number = {7},
  pages = {073901},
  year = {2007},
  month = {Aug},
  publisher = {American Physical Society},
  doi = {10.1103/PhysRevLett.99.073901}
}

@article{BERARD2006190,
  title = {Spin {Hall} effect and {Berry} phase of spinning particles},
  author = {B{\'e}rard, Alain and Mohrbach, Herv{\'e}},
  journal = {\href{https://doi.org/10.1016/j.physleta.2005.11.071}{Phys. Lett. A}},
  volume = {352},
  number = {3},
  pages = {190--195},
  year = {2006},
  doi = {10.1016/j.physleta.2005.11.071}
}

@article{BARONE2010267,
  title = {Transverse-spin and transverse-momentum effects in high-energy processes},
  author = {Barone, Vincenzo and Bradamante, Franco and Martin, Anna},
  journal = {\href{https://doi.org/10.1016/j.ppnp.2010.07.003}{Prog. Part. Nucl. Phys.}},
  volume = {65},
  number = {2},
  pages = {267--333},
  year = {2010},
  doi = {10.1016/j.ppnp.2010.07.003}
}

@article{santos2018laser,
  title = {Laser-driven strong magnetostatic fields with applications to charged beam transport and magnetized high energy-density physics},
  author = {Santos, Joao Jorge and Bailly-Grandvaux, M. and Ehret, M. and Arefiev, A. V. and Batani, D. and Beg, F. N. and Calisti, A. and Ferri, S. and Florido, R. and Forestier-Colleoni, P. and others},
  journal = {\href{https://doi.org/10.1063/1.5018735}{Phys. Plasmas}},
  volume = {25},
  number = {5},
  pages = {055701},
  year = {2018},
  publisher = {AIP Publishing},
  doi = {10.1063/1.5018735}
}

@article{cardano2015spin,
  title = {Spin--orbit photonics},
  author = {Cardano, Filippo and Marrucci, Lorenzo},
  journal = {\href{https://doi.org/10.1038/nphoton.2015.232}{Nat. Photonics}},
  volume = {9},
  number = {12},
  pages = {776--778},
  year = {2015},
  publisher = {Nature Publishing Group},
  doi = {10.1038/nphoton.2015.232}
}

@article{ShiDuYuan,
  title = {Spin photonics: from transverse spin to photonic skyrmions},
  author = {Shi, Peng and Du, Luping and Yuan, Xiaocong},
  journal = {\href{https://doi.org/10.1515/nanoph-2021-0046}{Nanophotonics}},
  volume = {10},
  number = {16},
  pages = {3927--3943},
  year = {2021},
  doi = {10.1515/nanoph-2021-0046}
}

@article{lu2014topological,
  title = {Topological photonics},
  author = {Lu, Ling and Joannopoulos, John D. and Solja{\v{c}}i{\'c}, Marin},
  journal = {\href{https://doi.org/10.1038/nphoton.2014.248}{Nat. Photonics}},
  volume = {8},
  number = {11},
  pages = {821--829},
  year = {2014},
  publisher = {Nature Publishing Group},
  doi = {10.1038/nphoton.2014.248}
}

@article{RevModPhys.91.015006,
  title = {Topological photonics},
  author = {Ozawa, Tomoki and Price, Hannah M. and Amo, Alberto and Goldman, Nathan and Hafezi, Mohammad and Lu, Ling and Rechtsman, Mikael C. and Schuster, David and Simon, Jonathan and Zilberberg, Oded and Carusotto, Iacopo},
  journal = {\href{https://doi.org/10.1103/RevModPhys.91.015006}{Rev. Mod. Phys.}},
  volume = {91},
  number = {1},
  pages = {015006},
  year = {2019},
  month = {Mar},
  publisher = {American Physical Society},
  doi = {10.1103/RevModPhys.91.015006}
}

@article{wang2009chiral,
  title = {Chiral metamaterials: simulations and experiments},
  author = {Wang, Bingnan and Zhou, Jiangfeng and Koschny, Thomas and Kafesaki, Maria and Soukoulis, Costas M.},
  journal = {\href{https://doi.org/10.1088/1464-4258/11/11/114003}{J. Opt. A: Pure Appl. Opt.}},
  volume = {11},
  number = {11},
  pages = {114003},
  year = {2009},
  publisher = {IOP Publishing},
  doi = {10.1088/1464-4258/11/11/114003}
}

@article{wang2016optical,
  title = {Optical chiral metamaterials: a review of the fundamentals, fabrication methods and applications},
  author = {Wang, Zuojia and Cheng, Feng and Winsor, Thomas and Liu, Yongmin},
  journal = {\href{https://doi.org/10.1088/0957-4484/27/41/412001}{Nanotechnology}},
  volume = {27},
  number = {41},
  pages = {412001},
  year = {2016},
  publisher = {IOP Publishing},
  doi = {10.1088/0957-4484/27/41/412001}
}

@misc{bhattacharjee2017topology,
  title = {Topology and Condensed Matter Physics},
  author = {Bhattacharjee, Somendra Mohan and Mj, Mahan and Bandyopadhyay, Abhijit},
  volume = {19},
  year = {2017},
  publisher = {Springer},
  doi = {10.1007/978-981-10-6841-6}
}

@misc{Shen2017TopologicalInsulators,
  author    = {Shun-Qing Shen},
  title     = {Topological Insulators: Dirac Equation in Condensed Matter},
  series    = {Springer Series in Solid-State Sciences},
  volume    = {187},
  edition   = {2},
  publisher = {Springer Singapore},
  year      = {2017},
  doi       = {10.1007/978-981-10-4606-3},
  isbn      = {978-981-10-4606-3}
}

@article{PhysRevLett.100.013904,
  title = {Possible Realization of Directional Optical Waveguides in Photonic Crystals with Broken Time-Reversal Symmetry},
  author = {Haldane, F. D. M. and Raghu, S.},
  journal = {\href{https://doi.org/10.1103/PhysRevLett.100.013904}{Phys. Rev. Lett.}},
  volume = {100},
  number = {1},
  pages = {013904},
  year = {2008},
  month = {Jan},
  publisher = {American Physical Society},
  doi = {10.1103/PhysRevLett.100.013904}
}

@article{PhysRevA.78.033834,
  title = {Analogs of quantum-{Hall}-effect edge states in photonic crystals},
  author = {Raghu, S. and Haldane, F. D. M.},
  journal = {\href{https://doi.org/10.1103/PhysRevA.78.033834}{Phys. Rev. A}},
  volume = {78},
  number = {3},
  pages = {033834},
  year = {2008},
  month = {Sep},
  publisher = {American Physical Society},
  doi = {10.1103/PhysRevA.78.033834}
}

@article{PhysRevLett.124.073603,
  title = {Topological Bands and Triply Degenerate Points in Non-Hermitian Hyperbolic Metamaterials},
  author = {Hou, Junpeng and Li, Zhitong and Luo, Xi-Wang and Gu, Qing and Zhang, Chuanwei},
  journal = {\href{https://doi.org/10.1103/PhysRevLett.124.073603}{Phys. Rev. Lett.}},
  volume = {124},
  number = {7},
  pages = {073603},
  year = {2020},
  month = {Feb},
  publisher = {American Physical Society},
  doi = {10.1103/PhysRevLett.124.073603}
}

@article{PhysRevLett.114.037402,
  title = {Topological Photonic Phase in Chiral Hyperbolic Metamaterials},
  author = {Gao, Wenlong and Lawrence, Mark and Yang, Biao and Liu, Fu and Fang, Fengzhou and B\'eri, Benjamin and Li, Jensen and Zhang, Shuang},
  journal = {\href{https://doi.org/10.1103/PhysRevLett.114.037402}{Phys. Rev. Lett.}},
  volume = {114},
  number = {3},
  pages = {037402},
  year = {2015},
  month = {Jan},
  publisher = {American Physical Society},
  doi = {10.1103/PhysRevLett.114.037402}
}

@article{PhysRevA.93.061801,
  title = {Topological photonic crystal with equifrequency {Weyl} points},
  author = {Wang, Luyang and Jian, Shao-Kai and Yao, Hong},
  journal = {\href{https://doi.org/10.1103/PhysRevA.93.061801}{Phys. Rev. A}},
  volume = {93},
  number = {6},
  pages = {061801},
  year = {2016},
  month = {Jun},
  publisher = {American Physical Society},
  doi = {10.1103/PhysRevA.93.061801}
}

@article{PhysRevLett.104.087401,
  title = {Photonic Band Structure of Dispersive Metamaterials Formulated as a Hermitian Eigenvalue Problem},
  author = {Raman, Aaswath and Fan, Shanhui},
  journal = {\href{https://doi.org/10.1103/PhysRevLett.104.087401}{Phys. Rev. Lett.}},
  volume = {104},
  number = {8},
  pages = {087401},
  year = {2010},
  month = {Feb},
  publisher = {American Physical Society},
  doi = {10.1103/PhysRevLett.104.087401}
}

@article{Khanikaev2017,
  title = {Two-dimensional topological photonics},
  author = {Khanikaev, Alexander B. and Shvets, Gennady},
  journal = {\href{https://doi.org/10.1038/s41566-017-0048-5}{Nat. Photonics}},
  volume = {11},
  number = {12},
  pages = {763--773},
  year = {2017},
  doi = {10.1038/s41566-017-0048-5}
}

@article{PhysRevLett.114.114301,
  title = {Topological Acoustics},
  author = {Yang, Zhaoju and Gao, Fei and Shi, Xihang and Lin, Xiao and Gao, Zhen and Chong, Yidong and Zhang, Baile},
  journal = {\href{https://doi.org/10.1103/PhysRevLett.114.114301}{Phys. Rev. Lett.}},
  volume = {114},
  number = {11},
  pages = {114301},
  year = {2015},
  month = {Mar},
  publisher = {American Physical Society},
  doi = {10.1103/PhysRevLett.114.114301}
}

@article{Huber2022,
  title = {Topological acoustics},
  author = {Huber, Sebastian D. and Al{\`u}, Andrea},
  journal = {\href{https://doi.org/10.1038/s41578-022-00465-6}{Nat. Rev. Mater.}},
  volume = {7},
  number = {6},
  pages = {446--464},
  year = {2022},
  publisher = {Nature Publishing Group},
  doi = {10.1038/s41578-022-00465-6}
}

@article{PhysRevResearch.3.023109,
  title = {{Bloch}-type photonic skyrmions in optical chiral multilayers},
  author = {Zhang, Qiang and Xie, Zhenwei and Du, Luping and Shi, Peng and Yuan, Xiaocong},
  journal = {\href{https://doi.org/10.1103/PhysRevResearch.3.023109}{Phys. Rev. Res.}},
  volume = {3},
  number = {2},
  pages = {023109},
  year = {2021},
  month = {May},
  publisher = {American Physical Society},
  doi = {10.1103/PhysRevResearch.3.023109}
}

@article{bliokh2015quantum,
  title = {Quantum spin {Hall} effect of light},
  author = {Bliokh, Konstantin Y. and Smirnova, Daria and Nori, Franco},
  journal = {\href{https://doi.org/10.1126/science.aaa9519}{Science}},
  volume = {348},
  number = {6242},
  pages = {1448--1451},
  year = {2015},
  publisher = {American Association for the Advancement of Science},
  doi = {10.1126/science.aaa9519}
}

@article{yang2025induced,
  title = {Induced {Berry} connection and photonic spin {Hall} effect in optical {Dirac} theory},
  author = {Yang, Lili and Feng, Longlong},
  journal = {\href{https://doi.org/10.1103/b3ls-krx1}{Phys. Rev. A}},
  volume = {112},
  number = {1},
  pages = {013515},
  year = {2025},
  publisher = {American Physical Society},
  doi = {10.1103/b3ls-krx1}
}

@article{rabiPhysRev.51.652,
  title = {Space Quantization in a Gyrating Magnetic Field},
  author = {Rabi, I. I.},
  journal = {\href{https://doi.org/10.1103/PhysRev.51.652}{Phys. Rev.}},
  volume = {51},
  number = {8},
  pages = {652--654},
  year = {1937},
  month = {Apr},
  publisher = {American Physical Society},
  doi = {10.1103/PhysRev.51.652}
}

@article{rabiobservationPhysRevLett.58.353,
  title = {Observation of quantum collapse and revival in a one-atom maser},
  author = {Rempe, Gerhard and Walther, Herbert and Klein, Norbert},
  journal = {\href{https://doi.org/10.1103/PhysRevLett.58.353}{Phys. Rev. Lett.}},
  volume = {58},
  number = {4},
  pages = {353--356},
  year = {1987},
  month = {Jan},
  publisher = {American Physical Society},
  doi = {10.1103/PhysRevLett.58.353}
}

@article{shandarova2009experimental,
  title = {Experimental observation of {Rabi} oscillations in photonic lattices},
  author = {Shandarova, Ksenia and R{\"u}ter, Christian E. and Kip, Detlef and Makris, Konstantinos G. and Christodoulides, Demetrios N. and Peleg, Or and Segev, Mordechai},
  journal = {\href{https://doi.org/10.1103/PhysRevLett.102.123905}{Phys. Rev. Lett.}},
  volume = {102},
  number = {12},
  pages = {123905},
  year = {2009},
  publisher = {American Physical Society},
  doi = {10.1103/PhysRevLett.102.123905}
}

@article{liu2023spin,
  title = {Spin-orbit {Rabi} oscillations in optically synthesized magnetic fields},
  author = {Liu, Guohua and Zhang, Xiliang and Zhang, Xin and Hu, Yanwen and Li, Zhen and Chen, Zhenqiang and Fu, Shenhe},
  journal = {\href{https://doi.org/10.1038/s41377-023-01238-8}{Light Sci. Appl.}},
  volume = {12},
  number = {1},
  pages = {205},
  year = {2023},
  publisher = {Nature Publishing Group},
  doi = {10.1038/s41377-023-01238-8}
}

@article{PhysRevLett.105.133603,
  title = {Observation of the {Rabi} Oscillation of Light Driven by an Atomic Spin Wave},
  author = {Chen, L. Q. and Zhang, Guo-Wan and Bian, Cheng-ling and Yuan, Chun-Hua and Ou, Z. Y. and Zhang, Weiping},
  journal = {\href{https://doi.org/10.1103/PhysRevLett.105.133603}{Phys. Rev. Lett.}},
  volume = {105},
  number = {13},
  pages = {133603},
  year = {2010},
  month = {Sep},
  publisher = {American Physical Society},
  doi = {10.1103/PhysRevLett.105.133603}
}

@article{PhysRevLett.125.123201,
  title = {Unveiling Chiral Phase Evolution in {Rabi} Oscillations from a Photonic Setting},
  author = {Zhang, Ping and Kang, Qianqian and Pei, Yumiao and Wang, Zhaoyuan and Hu, Yi and Chen, Zhigang and Xu, Jingjun},
  journal = {\href{https://doi.org/10.1103/PhysRevLett.125.123201}{Phys. Rev. Lett.}},
  volume = {125},
  number = {12},
  pages = {123201},
  year = {2020},
  month = {Sep},
  publisher = {American Physical Society},
  doi = {10.1103/PhysRevLett.125.123201}
}

@article{zhong2019rabi,
  title = {{Rabi}-like oscillation of photonic topological valley {Hall} edge states},
  author = {Zhong, Hua and Kartashov, Yaroslav V. and Zhang, Yiqi and Song, Daohong and Zhang, Yanpeng and Li, Fuli and Chen, Zhigang},
  journal = {\href{https://doi.org/10.1364/OL.44.003342}{Opt. Lett.}},
  volume = {44},
  number = {13},
  pages = {3342--3345},
  year = {2019},
  publisher = {Optica Publishing Group},
  doi = {10.1364/OL.44.003342}
}

@article{Stalder:96,
  title = {Linearly polarized light with axial symmetry generated by liquid-crystal polarization converters},
  author = {Stalder, M. and Schadt, M.},
  journal = {\href{https://doi.org/10.1364/OL.21.001948}{Opt. Lett.}},
  volume = {21},
  number = {23},
  pages = {1948--1950},
  year = {1996},
  month = {Dec},
  publisher = {Optica Publishing Group},
  doi = {10.1364/OL.21.001948}
}

@article{Bomzon:02,
  title = {Radially and azimuthally polarized beams generated by space-variant dielectric subwavelength gratings},
  author = {Bomzon, Ze'ev and Biener, Gabriel and Kleiner, Vladimir and Hasman, Erez},
  journal = {\href{https://doi.org/10.1364/OL.27.000285}{Opt. Lett.}},
  volume = {27},
  number = {5},
  pages = {285--287},
  year = {2002},
  month = {Mar},
  publisher = {Optica Publishing Group},
  doi = {10.1364/OL.27.000285}
}

@article{gorlach2018far,
  title = {Far-field probing of leaky topological states in all-dielectric metasurfaces},
  author = {Gorlach, Maxim A. and Ni, Xiang and Smirnova, Daria A. and Korobkin, Dmitry and Zhirihin, Dmitry and Slobozhanyuk, Alexey P. and Belov, Pavel A. and Al{\`u}, Andrea and Khanikaev, Alexander B.},
  journal = {\href{https://doi.org/10.1038/s41467-018-03330-9}{Nat. Commun.}},
  volume = {9},
  number = {1},
  pages = {909},
  year = {2018},
  publisher = {Nature Publishing Group},
  doi = {10.1038/s41467-018-03330-9}
}

@article{liang2025metasurface,
  title = {Metasurface-enabled quantum holograms with hybrid entanglement},
  author = {Liang, Hong and Wong, Wai Chun and An, Tailin and Li, Jensen},
  journal = {\href{https://doi.org/10.1117/1.AP.7.2.026006}{Adv. Photonics}},
  volume = {7},
  number = {2},
  pages = {026006},
  year = {2025},
  publisher = {Society of Photo-Optical Instrumentation Engineers},
  doi = {10.1117/1.AP.7.2.026006}
}

@article{alarcon2023all,
  title = {All-in-fiber dynamically reconfigurable orbital angular momentum mode sorting},
  author = {Alarc{\'o}n, Alvaro and G{\'o}mez, Santiago and Spegel-Lexne, Daniel and Argillander, Joakim and Cari{\~n}e, Jaime and Ca{\~n}as, Gustavo and Lima, Gustavo and Xavier, Guilherme B.},
  journal = {\href{https://doi.org/10.1021/acsphotonics.3c00825}{ACS Photonics}},
  volume = {10},
  number = {10},
  pages = {3700--3707},
  year = {2023},
  publisher = {ACS Publications},
  doi = {10.1021/acsphotonics.3c00825}
}

@article{qiu2020efficient,
  title = {Efficient all-optical router and beam splitter for light with orbital angular momentum},
  author = {Qiu, T. H. and Li, H. and Xie, M. and Liu, Q. and Ma, H. Y. and Xu, R.},
  journal = {\href{https://doi.org/10.1364/OE.395984}{Opt. Express}},
  volume = {28},
  number = {13},
  pages = {19750--19759},
  year = {2020},
  publisher = {Optica Publishing Group},
  doi = {10.1364/OE.395984}
}

@article{luo2017synthetic,
  title = {Synthetic-lattice enabled all-optical devices based on orbital angular momentum of light},
  author = {Luo, Xi-Wang and Zhou, Xingxiang and Xu, Jin-Shi and Li, Chuan-Feng and Guo, Guang-Can and Zhang, Chuanwei and Zhou, Zheng-Wei},
  journal = {\href{https://doi.org/10.1038/ncomms16097}{Nat. Commun.}},
  volume = {8},
  number = {1},
  pages = {16097},
  year = {2017},
  publisher = {Nature Publishing Group},
  doi = {10.1038/ncomms16097}
}

@article{forbes2021structured,
  title = {Structured light},
  author = {Forbes, Andrew and de Oliveira, Michael and Dennis, Mark R.},
  journal = {\href{https://doi.org/10.1038/s41566-021-00780-4}{Nat. Photonics}},
  volume = {15},
  number = {4},
  pages = {253--262},
  year = {2021},
  publisher = {Nature Publishing Group},
  doi = {10.1038/s41566-021-00780-4}
}

@article{shen2019optical,
  title = {Optical vortices 30 years on: {OAM} manipulation from topological charge to multiple singularities},
  author = {Shen, Yijie and Wang, Xuejiao and Xie, Zhenwei and Min, Changjun and Fu, Xing and Liu, Qiang and Gong, Mali and Yuan, Xiaocong},
  journal = {\href{https://doi.org/10.1038/s41377-019-0194-2}{Light Sci. Appl.}},
  volume = {8},
  number = {1},
  pages = {90},
  year = {2019},
  publisher = {Nature Publishing Group},
  doi = {10.1038/s41377-019-0194-2}
}

@article{liu2023higher,
  title = {Higher-order optical {Rabi} oscillations},
  author = {Liu, Guohua and Fu, Shenhe and Zhu, Siqi and Yin, Hao and Li, Zhen and Chen, Zhenqiang},
  journal = {\href{https://doi.org/10.1016/j.fmre.2022.03.014}{Fundam. Res.}},
  volume = {3},
  number = {6},
  pages = {898--903},
  year = {2023},
  publisher = {Elsevier},
  doi = {10.1016/j.fmre.2022.03.014}
}

@article{zhan2009cylindrical,
  title = {Cylindrical vector beams: from mathematical concepts to applications},
  author = {Zhan, Qiwen},
  journal = {\href{https://doi.org/10.1364/AOP.1.000001}{Adv. Opt. Photonics}},
  volume = {1},
  number = {1},
  pages = {1--57},
  year = {2009},
  publisher = {Optica Publishing Group},
  doi = {10.1364/AOP.1.000001}
}

@article{PhysRevB.88.085111,
  title = {Role of duality symmetry in transformation optics},
  author = {Fernandez-Corbaton, Ivan and Molina-Terriza, Gabriel},
  journal = {\href{https://doi.org/10.1103/PhysRevB.88.085111}{Phys. Rev. B}},
  volume = {88},
  number = {8},
  pages = {085111},
  year = {2013},
  month = {Aug},
  publisher = {American Physical Society},
  doi = {10.1103/PhysRevB.88.085111}
}

@article{khorasaninejad2017metalenses,
  title = {Metalenses: Versatile multifunctional photonic components},
  author = {Khorasaninejad, Mohammadreza and Capasso, Federico},
  journal = {\href{https://doi.org/10.1126/science.aam8100}{Science}},
  volume = {358},
  number = {6367},
  pages = {eaam8100},
  year = {2017},
  publisher = {American Association for the Advancement of Science},
  doi = {10.1126/science.aam8100}
}

@article{maguid2016photonic,
  title = {Photonic spin-controlled multifunctional shared-aperture antenna array},
  author = {Maguid, Elhanan and Yulevich, Igor and Veksler, Dekel and Kleiner, Vladimir and Brongersma, Mark L. and Hasman, Erez},
  journal = {\href{https://doi.org/10.1126/science.aaf3417}{Science}},
  volume = {352},
  number = {6290},
  pages = {1202--1206},
  year = {2016},
  publisher = {American Association for the Advancement of Science},
  doi = {10.1126/science.aaf3417}
}

@article{PhysRevA.109.023522,
  title = {Photonic spin-orbit coupling induced by deep-subwavelength structured light},
  author = {Zhang, Xin and Liu, Guohua and Hu, Yanwen and Lin, Haolin and Zeng, Zepei and Zhang, Xiliang and Li, Zhen and Chen, Zhenqiang and Fu, Shenhe},
  journal = {\href{https://doi.org/10.1103/PhysRevA.109.023522}{Phys. Rev. A}},
  volume = {109},
  number = {2},
  pages = {023522},
  year = {2024},
  month = {Feb},
  publisher = {American Physical Society},
  doi = {10.1103/PhysRevA.109.023522}
}

@misc{joannopoulos2008photonic,
  title = {Photonic Crystals: Molding the Flow of Light},
  author = {Joannopoulos, John D. and Johnson, Steven G. and Winn, Joshua N. and Meade, Robert D.},
  edition = {2nd},
  year = {2008},
  publisher = {Princeton University Press},
  address = {Princeton, NJ},
  isbn = {9780691124568},
  doi={10.1515/9781400828241}
}

@article{zhang2024metasurface,
  title={Metasurface-tuned light-matter interactions for high-performance photodetectors},
  author={Zhang, Guanyu and Xu, Chang and Sun, Dong and Wang, Qinsheng and Lu, Guowei and Gong, Qihuang},
  journal={\href{https://doi.org/10.1016/j.fmre.2024.01.002}{Fundam. Res.}},
  year={2024},
  publisher={Elsevier},
  doi={10.1016/j.fmre.2024.01.002}
}

@article{shi2021quantum,
  title={Quantum interface for high-dimensional quantum states encoded in an orbital angular momentum space},
  author={Shi, Bao-Sen and Zhou, Zhi-Yuan},
  journal={\href{https://doi.org/10.1016/j.fmre.2021.01.001}{Fundam. Res.}},
  volume={1},
  number={1},
  pages={88--90},
  year={2021},
  publisher={Elsevier},
  doi={10.1016/j.fmre.2021.01.001}
}

@article{chen2023spin,
  title={Spin vector potential and spin Aharonov-Bohm effect},
  author={Chen, Jing-Ling and Fan, Xing-Yan and Xie, Xiang-Ru},
  journal={\href{https://doi.org/10.1016/j.fmre.2023.05.021}{Fundam. Res.}},
  year={2023},
  publisher={Elsevier},
  doi={10.1016/j.fmre.2023.05.021}
}

@article{huang2024observation,
  title={Observation of High-Order Quantum Pancharatnam-Berry Phase with Structured Photons},
  author={Huang, Shuang-Yin and Jiang, He and Ren, Zhi-Cheng and Cheng, Zi-Mo and Zhu, Wen-Zheng and Gao, Jing and Liu, Chang and Wang, Xi-Lin and Wang, Hui-Tian},
  journal={\href{https://doi.org/10.1016/j.fmre.2024.11.016}{Fundam. Res.}},
  year={2024},
  publisher={Elsevier},
  doi={10.1016/j.fmre.2024.11.016}
}

@article{makris2008optical, 
    title={Optical transitions and Rabi oscillations in waveguide arrays}, 
    author={Makris, KG and Christodoulides, DN and Peleg, O and Segev, M and Kip, Detlef}, journal={\href{https://doi.org/10.1364/OE.16.010309}{Opt. Express}}, 
    volume={16}, 
    number={14}, 
    pages={10309--10314}, 
    year={2008}, 
    publisher={Optical Society of America}, 
    doi={10.1364/OE.16.010309} 
}

@article{zhang2015rabi,
  title={Rabi oscillations and stimulated mode conversion on the subwavelength scale},
  author={Zhang, Xiao and Ye, Fangwei and Kartashov, Yaroslav V and Chen, Xianfeng},
  journal={\href{https://doi.org/10.1364/OE.23.006731}{Opt. Express}},
  volume={23},
  number={5},
  pages={6731--6737},
  year={2015},
  publisher={Optical Society of America},
  doi={10.1364/OE.23.006731}
}

@article{jin2020rabi,
  title={Rabi oscillations of azimuthons in weakly nonlinear waveguides},
  author={Jin, Kaichao and Li, Yongdong and Li, Feng and Beli{\'c}, Milivoj R and Zhang, Yanpeng and Zhang, Yiqi},
  journal={\href{https://doi.org/10.1117/1.AP.2.4.046002}{Adv. Photonics}},
  volume={2},
  number={4},
  pages={046002--046002},
  year={2020},
  publisher={Society of Photo-Optical Instrumentation Engineers},
  doi={10.1117/1.AP.2.4.046002}
}

@article{vijay2012stabilizing,
  title={Stabilizing Rabi oscillations in a superconducting qubit using quantum feedback},
  author={Vijay, R and Macklin, Chris and Slichter, DH and Weber, SJ and Murch, KW and Naik, Ravi and Korotkov, Alexander N and Siddiqi, Irfan},
  journal={\href{https://doi.org/10.1038/nature11505}{Nature}},
  volume={490},
  number={7418},
  pages={77--80},
  year={2012},
  publisher={Nature Publishing Group UK London},
  doi={10.1038/nature11505}
}

@article{cronenberg2018acoustic,
  title={Acoustic Rabi oscillations between gravitational quantum states and impact on symmetron dark energy},
  author={Cronenberg, Gunther and Brax, Philippe and Filter, Hanno and Geltenbort, Peter and Jenke, Tobias and Pignol, Guillaume and Pitschmann, Mario and Thalhammer, Martin and Abele, Hartmut},
  journal={\href{https://doi.org/10.1038/s41567-018-0205-x}{Nat. Phys.}},
  volume={14},
  number={10},
  pages={1022--1026},
  year={2018},
  publisher={Nature Publishing Group UK London},
  doi={10.1038/s41567-018-0205-x}
}

@article{PhysRevLett.102.123905,
  title = {Experimental Observation of Rabi Oscillations in Photonic Lattices},
  author = {Shandarova, Ksenia and R\"uter, Christian E. and Kip, Detlef and Makris, Konstantinos G. and Christodoulides, Demetrios N. and Peleg, Or and Segev, Mordechai},
  journal = {\href{https://doi.org/10.1103/PhysRevLett.102.123905}{Phys. Rev. Lett.}},
  volume = {102},
  issue = {12},
  pages = {123905},
  numpages = {4},
  year = {2009},
  month = {Mar},
  publisher = {American Physical Society},
  doi = {10.1103/PhysRevLett.102.123905}
}

@article{kitano2017rabi,
  title={Rabi oscillations in the spatial profiles of superfluorescent pulses from rubidium vapor},
  author={Kitano, K and Maeda, H},
  journal={\href{https://doi.org/10.1364/OE.25.023826}{Opt. Express}},
  volume={25},
  number={20},
  pages={23826--23832},
  year={2017},
  publisher={Optical Society of America},
  doi={10.1364/OE.25.023826}
}

@article{k2th-m73q,
  title = {Photonic Rabi Oscillations in Defective Plasma Photonic Crystals},
  author = {Zhang, Xiao-Bo and Weng, Su-Ming and Ai, Hong and Qiao, Xin and Xue, Ju-Kui and Sheng, Zheng-Ming},
  journal = {\href{https://doi.org/10.1103/k2th-m73q}{Phys. Rev. Lett.}},
  volume = {135},
  issue = {1},
  pages = {015101},
  numpages = {5},
  year = {2025},
  month = {Jul},
  publisher = {American Physical Society},
  doi = {10.1103/k2th-m73q}
}

@article{KELLER20051,
title = {On the theory of spatial localization of photons},
journal = {\href{https://doi.org/10.1016/j.physrep.2005.01.002}{Phys. Rep.}},
volume = {411},
number = {1},
pages = {1-232},
year = {2005},
issn = {0370-1573},
doi = {https://doi.org/10.1016/j.physrep.2005.01.002},
author = {Ole Keller}
}

@article{YANG2026170557,
title = {Photonic Kramers-like degeneracy and spin-orbital Hall effects of light},
journal = {\href{https://doi.org/10.1016/j.aop.2026.170557}{Ann. Phys.}},
volume = {492},
pages = {170557},
year = {2026},
issn = {0003-4916},
doi = {https://doi.org/10.1016/j.aop.2026.170557},
author = {Lili Yang and Pengming Zhang and Yonghui Qi and Longlong Feng}
}

@article{Zheng_2021,
doi = {10.1209/0295-5075/ac3e8b},
year = {2022},
month = {feb},
publisher = {EDP Sciences, IOP Publishing and Società Italiana di Fisica},
volume = {136},
number = {3},
pages = {30002},
author = {Zheng, Chao},
title = {Quantum simulation of PT-arbitrary-phase–symmetric systems},
journal = {\href{https://doi.org/10.1209/0295-5075/ac3e8b}{Europhys. Lett.}}
}

@misc{bender2018pt,
  title={PT Symmetry: In Quantum and Classical Physics},
  author={Bender, C.M. and Dorey, P.E. and Dunning, C. and Fring, A. and Hook, D.W. and Jones, H.F. and Kuzhel, S. and L{\'e}vai, G. and Tateo, R.},
  isbn={9781786345974},
  doi={https://doi.org/10.1142/q0178},
  year={2018},
  publisher={World Scientific Publishing Company}
}

\begin{appendix}
\section{Derivation of the Dispersive Optical Eigenvalue Equation in Chiral Media}\label{appA}

In frequency space at angular frequency $\omega$, the displacement $\mathbf{D}(\mathbf{r})$ and magnetic induction $\mathbf{B}(\mathbf{r})$ are related to $\mathbf{E}(\mathbf{r})$ and $\mathbf{H}(\mathbf{r})$ via the dispersive constitutive relations:
\begin{equation}
\begin{aligned}
\mathbf{D}(\mathbf{r}) &= \boldsymbol{\epsilon}\cdot \mathbf{E}(\mathbf{r}) + i\boldsymbol{\xi}\cdot \mathbf{H}(\mathbf{r}), \\
\mathbf{B}(\mathbf{r}) &= \boldsymbol{\mu}\cdot \mathbf{H}(\mathbf{r}) - i\boldsymbol{\zeta}\cdot \mathbf{E}(\mathbf{r}),
\end{aligned}
\end{equation}
where $\boldsymbol{\epsilon}$ and $\boldsymbol{\mu}$ are the dispersive permittivity and permeability tensors, and $\boldsymbol{\xi}$ and $\boldsymbol{\zeta}$ denote the magnetoelectric coupling tensors, satisfying the reciprocity condition $\boldsymbol{\zeta} = \boldsymbol{\xi}^T$.

The inverse dispersive constitutive relations read
\begin{equation}
\begin{aligned}
\mathbf{E}(\mathbf{r}) &= \boldsymbol{\varphi}\cdot \mathbf{D}(\mathbf{r}) - i\boldsymbol{\beta}\cdot \mathbf{B}(\mathbf{r}), \\
\mathbf{H}(\mathbf{r}) &= \boldsymbol{\vartheta}\cdot \mathbf{B}(\mathbf{r}) + i\boldsymbol{\alpha}\cdot \mathbf{D}(\mathbf{r}),
\end{aligned}
\end{equation}
where
\begin{equation}
\begin{aligned}
\boldsymbol{\varphi} &= \left[\boldsymbol{\epsilon} - \boldsymbol{\xi}\boldsymbol{\mu}^{-1}\boldsymbol{\zeta}\right]^{-1}, \quad &\boldsymbol{\beta} &= \boldsymbol{\varphi}\cdot\boldsymbol{\xi}\cdot\boldsymbol{\mu}^{-1}, \\ 
\boldsymbol{\vartheta} &= \left[\boldsymbol{\mu} - \boldsymbol{\zeta}\boldsymbol{\epsilon}^{-1}\boldsymbol{\xi}\right]^{-1}, \quad &\boldsymbol{\alpha} &= \boldsymbol{\vartheta}\cdot\boldsymbol{\zeta}\cdot\boldsymbol{\epsilon}^{-1}.
\end{aligned}
\end{equation}

In terms of these inverse tensors, Maxwell's curl equations in frequency space become
\begin{equation}\label{eq:MaxwellEq_in_chiralmedia}
\begin{aligned}
	-i\omega \mathbf{D}(\mathbf{r}) &= \nabla \times\left[\boldsymbol{\vartheta} \cdot \mathbf{B}(\mathbf{r}) + i\boldsymbol{\alpha} \cdot \mathbf{D}(\mathbf{r})\right], \\
	i\omega \mathbf{B}(\mathbf{r}) &= -\nabla \times\left[\boldsymbol{\varphi} \cdot \mathbf{D}(\mathbf{r}) - i\boldsymbol{\beta} \cdot \mathbf{B}(\mathbf{r})\right].
\end{aligned}
\end{equation}

We follow the approach of Feng \& Wu to derive the optical eigenvalue equation in chiral media. Our analysis is carried out in the helicity basis spanned by the transverse basis vectors ${\bf e}_{\pm}$ with respect to a given direction, which may be chosen to coincide with the propagation direction of the incident light. Without loss of generality, we take the propagation direction to be along the $z$-axis, with unit vector ${\bf e}_z$. The helicity basis ${\bf e}_{\pm}$ satisfies
$({\bf e}_z\cdot{\bf s}){\bf e}_{\pm} = \pm {\bf e}_{\pm}$,
and is related to the Cartesian basis by
${\bf e}_{\pm} = ({\bf e}_x\pm i{\bf e}_y)/\sqrt{2}$.
Accordingly, for a given vector ${\bf V}$, its Cartesian components are related to its components in helicity space by the unitary transformation $(V_x, V_y, V_z)^T = \hat{U}(V_+, V_-, V_z)^T$, where
\begin{equation}
\hat{U} = \frac{1}{\sqrt{2}}\begin{pmatrix} 1 & 1 & 0 \\ i & -i & 0 \\ 0 & 0 & \sqrt{2} \end{pmatrix}.
\end{equation}

Each curl operation in Eq.~(\ref{eq:MaxwellEq_in_chiralmedia}) can be expressed as an operator acting on field vector $\mathbf{V}$:
\begin{equation}\label{eq:operator_Hxcurl}
\hat{\mathcal{H}} \circ \mathbf{V} = \nabla \times \left[\Pi \circ \mathbf{V}\right] = \left[\left(\hat{\mathbf{k}}\cdot\mathbf{s}\right) \cdot \Pi\right] \circ \mathbf{V},
\end{equation}
where $\Pi$ represents any rank-2 dispersive inverse tensor, $\hat{\mathbf{k}} = -i\nabla$ is the momentum operator, and $\mathbf{s}$ denotes the spin-1 vector operator with adjoint components $(s_i)_{jk} = -i\epsilon_{ijk}$.

Transforming to helicity space, the operator $\hat{\mathcal{H}}$ expands as
\begin{equation}\label{eq:H-transformation}
\begin{aligned}
\hat{U}^{-1} \hat{\mathcal{H}} \hat{U} &= \left(\hat{U}^{-1} (\hat{\mathbf{k}}\cdot\mathbf{s}) \hat{U}\right) \cdot \left(\hat{U}^{-1} \Pi \hat{U}\right) \\
&= \begin{pmatrix} \hat{k}_z \sigma_3 & -\sigma_3\hat{\mathbf{k}}_\perp \\ -(\sigma_3\hat{\mathbf{k}}_\perp)^\dagger & 0 \end{pmatrix} \cdot \begin{pmatrix} Q_0 \mathbf{I} + \mathbf{Q}\cdot\boldsymbol{\sigma} & \mathbf{q} \\ \mathbf{q}^\dagger & q_0 \end{pmatrix} \\
&= \begin{pmatrix} \sigma_3\left[\hat{k}_z\left(Q_0\sigma_0 + \mathbf{Q}\cdot\boldsymbol{\sigma}\right) - \hat{\mathbf{k}}_\perp \mathbf{q}^\dagger\right] & \sigma_3\left(\hat{k}_z\mathbf{q} - \hat{\mathbf{k}}_\perp q_0\right) \\ -(\sigma_3 \hat{\mathbf{k}}_\perp^\dagger)\left(Q_0\sigma_0 + \mathbf{Q}\cdot\boldsymbol{\sigma}\right) & -(\sigma_3 \hat{\mathbf{k}}_\perp^\dagger)\mathbf{q} \end{pmatrix},
\end{aligned}
\end{equation}
where $\hat{\mathbf{k}}_\perp = (\hat{k}_+, \hat{k}_-)^T$ is the transverse momentum vector with ladder components $\hat{k}_\pm = (\hat{k}_x \mp i\hat{k}_y)/\sqrt{2}$, $\sigma_0$ is the $2\times 2$ identity matrix, $\boldsymbol{\sigma} = (\boldsymbol{\sigma}_\perp, \sigma_3)$ are the Pauli matrices, and the frequency-dependent multipole moments of tensor $\Pi$ are defined as
\begin{equation}
\begin{aligned}
   Q_0 &= \frac{1}{2}\left[\Pi_{11}+\Pi_{22}\right], \quad &q_0 &= \Pi_{33}, \\
   Q_{\pm 1} &= \frac{1}{\sqrt{2}}\left[\Pi_{13}\mp i\Pi_{23}\right], \quad &Q_z &= \frac{i}{2}\left[\Pi_{12}-\Pi_{21}\right], \\
   Q_{\pm 2} &= \frac{1}{2}\left[\Pi_{11} - \Pi_{22}\right] \mp \frac{i}{2}\left[\Pi_{12}+\Pi_{21}\right].
\end{aligned}
\end{equation}

The divergence-free transversality condition $\nabla\cdot\mathbf{V}(\mathbf{r}) = 0$ constrains the longitudinal component:
\begin{equation}\label{eq:z-transversality}
V_z(\mathbf{r}) = -\hat{k}_z^{-1}\hat{\mathbf{k}}_\perp^\dagger \mathbf{V}_\perp(\mathbf{r}),
\end{equation}
allowing exact reduction of the 3-vector field to its transverse components $\mathbf{V}_\perp = (V_+, V_-)^T$. Substituting Eq.~(\ref{eq:z-transversality}) into Eq.~(\ref{eq:H-transformation}) yields the effective $2\times 2$ operator acting strictly on transverse helicity space:
\begin{equation}\label{eq:projectedHxcurl}
\hat{\mathcal{H}}_\perp\mathbf{V}_\perp = \left[\hat{\mathcal{H}}\mathbf{V}\right]_\perp = \sigma_3 \left[ \mathcal{H}_0\sigma_0 + \boldsymbol{\mathcal{H}}_\perp \cdot \boldsymbol{\sigma}_\perp + \mathcal{H}_3\sigma_3 \right]\mathbf{V}_\perp,
\end{equation}
where the dispersive operator components expand as
\begin{equation}\label{eq:effH_components}
\begin{aligned}
\hat{\mathcal{H}}_0 &= \hat{k}_z Q_0 + q_0\frac{1}{\hat{k}_z} \hat{k}_+ \hat{k}_- - 2\mathbf{q}^\dagger \cdot \hat{\mathbf{k}}_\perp, \\
\hat{\boldsymbol{\mathcal{H}}}_\perp &= \hat{k}_z \mathbf{Q}_\perp + q_0\frac{1}{\hat{k}_z} \hat{\mathbf{k}}_\perp^t - 2\hat{\mathbf{k}}_\perp^q, \\
\hat{\mathcal{H}}_3 &= Q_z,
\end{aligned}
\end{equation}
with vector operators $\hat{\mathbf{k}}_\perp^q = \left(Q_{+1}\hat{k}_+, Q_{-1}\hat{k}_-\right)^T$ and $\hat{\mathbf{k}}_\perp^t = \left(\hat{k}_+^2, \hat{k}_-^2\right)^T$.

Attaching a superscript $\Pi$ to distinguish corresponding material tensors, we define shorthand notations $H_0^\Pi = \hat{\mathcal{H}}_0^\Pi$ and $\mathbf{H}^\Pi = \left(\hat{\boldsymbol{\mathcal{H}}}_\perp^\Pi, \eta\hat{\mathcal{H}}_3^\Pi\right)$, where $\eta=1$ for electric tensors $\boldsymbol{\varphi}, \boldsymbol{\alpha}$ acting on $\mathbf{D}$, and $\eta=-1$ for magnetic tensors $\boldsymbol{\vartheta}, \boldsymbol{\beta}$ acting on $\mathbf{B}$.

Applying this transverse projection to Eq.~(\ref{eq:MaxwellEq_in_chiralmedia}), Maxwell's equations in helicity space read
\begin{equation}\label{eq:transverse_Maxwelleqn}
\begin{aligned}
\omega \mathbf{D}_\perp &= \left[H_0^\vartheta - \mathbf{H}^\vartheta \cdot \boldsymbol{\sigma}\right] i\sigma_3 \mathbf{B}_\perp - \sigma_3\left[H_0^\alpha + \mathbf{H}^\alpha \cdot \boldsymbol{\sigma}\right] \mathbf{D}_\perp, \\
\omega (i\sigma_3 \mathbf{B}_\perp) &= \left[H_0^\varphi + \mathbf{H}^\varphi \cdot \boldsymbol{\sigma}\right] \mathbf{D}_\perp - \sigma_3\left[H_0^\beta - \mathbf{H}^\beta \cdot \boldsymbol{\sigma}\right] i\sigma_3 \mathbf{B}_\perp.
\end{aligned}
\end{equation}

Introducing the frequency-dependent effective mass and momentum operators:
\begin{equation}\label{eq:def_mass_momentum}
\begin{aligned}
\hat{\mathrm{m}}_\pm &= \frac{1}{2}\left[H_0^\varphi \pm H_0^\vartheta\right], \quad &\hat{\mathrm{m}}_c &= \frac{1}{2}\left[H_0^\alpha \pm H_0^\beta\right], \\
\hat{\mathbf{p}}_\pm &= \frac{1}{2}\left[\mathbf{H}^\varphi \pm \mathbf{H}^\vartheta\right], \quad &\hat{\mathbf{p}}_c &= \frac{1}{2}\left[\mathbf{H}^\alpha \pm \mathbf{H}^\beta\right],
\end{aligned}
\end{equation}
and adopting the four-component bispinor $\boldsymbol{\Psi}_\perp = (\mathbf{D}_\perp + i\sigma_3\mathbf{B}_\perp, \mathbf{D}_\perp - i\sigma_3\mathbf{B}_\perp)^T$, Eq.~(\ref{eq:transverse_Maxwelleqn}) condenses into the exact frequency-domain optical Dirac-like eigenvalue equation:
\begin{equation}\label{eq:OpticalDiracEqn}
\begin{aligned}
\omega \boldsymbol{\Psi}_\perp(\mathbf{r}) = &\left[ \gamma_0\left(\hat{\mathrm{m}}_+ + \gamma_5 \hat{\mathrm{m}}_- - \gamma_3\left[\gamma_5\hat{\mathrm{m}}_+^c + \hat{\mathrm{m}}_-^c\right]\right) \right. \\
&\left. + \boldsymbol{\gamma}\cdot\left(\hat{\mathbf{p}}_+ + \gamma_5 \hat{\mathbf{p}}_- - \gamma_3\left[\gamma_5\hat{\mathbf{p}}_+^c + \hat{\mathbf{p}}_-^c\right]\right) \right] \boldsymbol{\Psi}_\perp(\mathbf{r}),
\end{aligned}
\end{equation}
where the standard Dirac representation of $\gamma$-matrices are used. Equation~(\ref{eq:OpticalDiracEqn}) corresponds strictly to the dispersive optical Hamiltonian $\hat{H}$ presented in Eq.~(\ref{eq:ODE-in-chiral}) of the main text.

\section{Validity of the paraxial reduction}\label{sec:AppB}

The derivation of Eq.~(\ref{eq:spatial-spinor-Hamiltonian}) assumes that the transverse momentum remains small compared with the longitudinal propagation constant of each helicity branch. More precisely, for a transverse Fourier component with \(q_\perp=|\mathbf q_\perp|\), the geometric paraxial parameter is
\begin{equation}
\epsilon_{\rm p}^{(\pm)}
=
\frac{q_\perp}{|k_{z}^{0,\pm}|}
\ll 1 ,
\end{equation}
where \(k_z^{0,\pm}=k_z^0\mp\Delta_\chi\) are the two chirality-split longitudinal propagation constants introduced above. For an \(LG_{p,\ell}\) mode with beam waist \(w_0\), the characteristic transverse momentum may be estimated as
\begin{equation}
q_{\perp,\mathrm{rms}}
\sim
\frac{\sqrt{2p+|\ell|+1}}{w_0}.
\end{equation}
Hence, higher-order radial or orbital-angular-momentum modes generally impose a more restrictive paraxial condition than the fundamental Gaussian mode. In particular, \(p=0\) and \(\ell=0\) correspond to the fundamental Gaussian beam, for which \(q_{\perp,\mathrm{rms}}\sim 1/w_0\).

This condition ensures that the longitudinal field components remain perturbative and that their feedback into the transverse dynamics first enters at second order in \(q_\perp/k_z\). Indeed, the divergence constraints \(\mathbf k\!\cdot\!\mathbf D=0\) and \(\mathbf k\!\cdot\!\mathbf B=0\) imply schematically
\begin{equation}
\begin{pmatrix}
E_z\\ H_z
\end{pmatrix}
=
-\frac{1}{k_z}
\mathbf C_z^{-1}
\mathbf Q_\perp
\mathbf C_t
\begin{pmatrix}
\mathbf E_\perp\\ \mathbf H_\perp
\end{pmatrix},
\end{equation}
where
\begin{equation}
\mathbf C_s=
\begin{pmatrix}
\epsilon_s&i\xi_s\\
-i\xi_s&\mu_s
\end{pmatrix},
\qquad s=t,z,
\end{equation}
and \(\mathbf Q_\perp\) denotes the projection along the transverse wavevector. Hence a convenient sufficient condition for the longitudinal sector to remain perturbative is
\begin{equation}
\frac{q_\perp}{|k_z|}
\left\|\mathbf C_z^{-1}\mathbf C_t\right\|
\ll1 .
\end{equation}
Since \(\xi_z=\chi_z\sqrt{\epsilon_z\mu_z}\), the longitudinal constitutive determinant is \(\det\mathbf C_z
=\epsilon_z\mu_z-\xi_z^2=\epsilon_z\mu_z(1-\chi_z^2).\)
Therefore the paraxial reduction additionally requires the system to remain sufficiently far from the singular constitutive limit \(|\chi_z|=1\), where \(\mathbf C_z^{-1}\) becomes strongly enhanced and even a small transverse momentum may generate appreciable longitudinal fields. Upon reinsertion of \(E_z\) and \(H_z\) into the transverse Maxwell equations, an additional transverse derivative is generated, so that the correction to the transverse propagation operator starts at \(O(q_\perp^2/k_z^2)\). Equation ~(\ref{eq:spatial-spinor-Hamiltonian}) should therefore be understood as a second-order paraxial Hamiltonian: the diffraction terms and the spin-orbit terms proportional to \(\hat k_\pm^2\) are retained consistently at \(O(q_\perp^2)\), whereas higher-order nonparaxial corrections are neglected.

Equivalently, at the level of the effective Hamiltonian Eq.~(\ref{eq:spatial-spinor-Hamiltonian}), the retained transverse correction must remain small compared with the leading longitudinal propagation scale. A conservative criterion is therefore
\begin{equation}
\frac{q_\perp^2}{2(k_z^0)^2}
\max\!\left\{
|\alpha_+|,\,
|\alpha_-|,\,
|\beta|
\right\}\ll1 .
\end{equation}
This form makes explicit that strong anisotropy or a near-singular longitudinal chiral response can reduce the range of transverse momenta over which the paraxial description is quantitatively reliable.

\end{appendix}

\end{multicols}
\end{document}